\documentclass[aps,pra,showpacs,twocolumn,a4paper,reprint,noeprint]{revtex4-1}
\usepackage[utf8]{inputenc}
\usepackage{graphicx}
\usepackage{amsmath}
\usepackage{amsfonts}
\usepackage{amssymb}
\usepackage[usenames,dvipsnames,svgnames,table]{xcolor}
\usepackage[USenglish]{babel}
\usepackage{hyperref}
\usepackage{grffile}

\newcommand{\ba}{\begin{eqnarray}}
\newcommand{\be}{\begin{equation}}
\newcommand{\ea}{\end{eqnarray}}
\newcommand{\ee}{\end{equation}}

\newcommand{\bra}[1]{ \left< #1 \right|}
\newcommand{\ket}[1]{ \left| #1 \right>}

\newcommand{\ketbra}[2]{ \ket{#1}\hspace{-0.3em}\bra{#2} }
\newcommand{\erwart}[3]{ \left< #1 \left| #2 \right| #3 \right> }
\newcommand{\expect}[1]{ \left< #1 \right> }

\DeclareMathOperator{\Tr}{Tr}

\newcommand{\affiliationTUD}{\affiliation{Institute of Theoretical Physics,
Technische Universit\"at Dresden, 01062 Dresden, Germany}}

\begin{document}

\title{Charge order in an interacting monolayer under transverse bias}

\author{Tim Ludwig}
\affiliationTUD

\author{Carsten Timm}
\affiliationTUD

\date{September 23, 2016}

\begin{abstract}
A monolayer of molecules or quantum dots sandwiched between electrodes
can be driven out of equilibrium by the application of a bias voltage between
the electrodes. We study charge ordering, i.e., the spontaneous formation
of a charge density wave, and the perpendicular current in
such a system within a master-equation approach augmented by mean-field
and classical Monte Carlo methods. Our approach is suitable for
weak tunneling between the monolayer and the electrodes.
For a square lattice with nearest-neighbor Coulomb repulsion, we present a
comprehensive study of the zero-temperature phases controlled by the on-site
energy, the bias voltage, and the degeneracy of the occupied single-site state.
One of the most interesting results is the prediction of a conducting
charge-density-wave phase that only occurs at a finite bias voltage.
We also study the universality classes of the phase transitions towards
charge-ordered states at zero and nonzero temperatures. While all
transitions at $T>0$ and some at $T=0$ belong to the two-dimensional Ising
universality class, we also find an absorbing-to-active phase transition in the
$\mathbb{Z}_2$ symmetric directed percolation (DP2) class at $T=0$.
\end{abstract}



\maketitle

\section{Introduction}
\label{sec.Introduction}

Layers of quantum dots or single molecules sandwiched between conducting
electrodes form promising systems for applications as well as for fundamental
research. For the past 20 years, experimentalists have investigated the
perpendicular current through self-organized layers of quantum dots
in semiconductor systems \cite{PhysRevB.54.16401, 
:/content/aip/journal/apl/70/1/10.1063/1.119276,
1347-4065-36-3S-1917, 0268-1242-14-11-104, belyaev:2001:ICPS25,
hapke-wurst_tuning_2003, lin_tunneling_2003, PhysRevB.68.155315,
Austing2005482, Sun01072006, PhysRevB.75.115315,
doi:10.1021/acs.chemrev.5b00678}.
These quantum-dot arrays were strongly disordered, though.
Molecular layers offer at least two advantages:
certain molecules readily form highly ordered self-assembled monolayers on
semiconducting or metallic substrates \cite{bigelow_oleophobic_1946, Scheffler201355,
C4NR06371F, doi:10.1021/la00022a004, :/content/aip/journal/apl/69/11/10.1063/1.117444,
Whitesides16042002} and individual molecules are, in principle, identical. 
On the other hand, it has proven to be difficult to fabricate a reliable
second (top) contact. Novel methods, such as low-energy,  indirect-path thermal
evaporation \cite{:/content/aip/journal/apl/85/14/10.1063/1.1799235},
rolled-up nanomembranes \cite{Prinz2000828,nature_410_168, doi:10.1021/jp409617r,
doi:10.1021/nl1010367, *doi:10.1021/nl1022145, *doi:10.1021/nl201773d,
*:/content/aip/journal/apl/100/2/10.1063/1.3676269, *doi:10.1021/nl303887b}
lift-off--float-on techniques \cite{nature_404_166, ADFM:ADFM_200290009,
doi:10.1021/jp104130w}, nanotransfer printing \cite{doi:10.1021/nl034217d,
doi:10.1021/ja901646j},
and transfer of multilayer graphene \cite{ADMA:ADMA201003178}
have been used to create reasonably homogeneous top contacts to molecular
layers. Controlled contacts are a prerequisite for the application of molecular
monolayers in electronic devices. Such applications are driven, on the one hand,
by the trend to further miniaturization, and, on the other, by the possibility
to functionalize the molecules~\cite{Vuillaume_ISBN9780199533060}.

Layers of molecules or quantum dots in sandwich structures also constitute
model systems for non-equilibrium statistical physics: a bias voltage applied
to the electrodes drives the system out of equilibrium. For time-independent
bias, the system approaches a stationary state, which is
characterized by a stationary current in the direction perpendicular to the
monolayer. For non-interacting quantum dots or molecules, the individual
entities conduct independently and the theoretical description can fall back on
transport theory for single dots or molecules \cite{2010Nanot..21A2001A, bruus2004many}.
The case of interacting dots or molecules  is much more interesting in
that it combines interactions with driving. Such a system can show
spontaneous symmetry breaking, begging the questions whether the
corresponding phase transitions belong to a universality class that is known
from equilibrium physics or to a genuinely non-equilibrium 
one~\cite{RevModPhys.49.435, RevModPhys.65.851,
2000AdPhy..49..815H, RevModPhys.76.663}.

In this paper, we study a relatively simple model in this class,
namely, a square array of sites that are either empty or singly occupied due
to a high charging energy and that interact through nearest-neighbor
Coulomb repulsion. We allow for arbitrary (spin or orbital) degeneracy of the
occupied states. The system is sandwiched between electrodes under an applied
bias voltage. As we shall see, this non-equilibrium situation can induce
spontaneous breaking of translational symmetry \cite{RevModPhys.65.851}
by the formation of a charge density wave with ordering
vector $(\pi,\pi)$, in which the two checkerboard sublattices have different
average occupation. We employ a mean-field
approximation in the framework of rate equations to obtain an overview of the
possible phases, and classical non-equilibrium Monte Carlo simulations
as an unbiased method to study them in more detail.

Our model is similar to the one studied by Kie\ss{}lich \textit{et al}.\
\cite{Kiesslich2002459, *PhysRevB.68.125331}
and by Wetzler \textit{et al}.\ \cite{PhysRevB.68.045323, *1367-2630-6-1-081},
but their systems were relatively small or
disordered. Their focus was on small arrays of quantum dots in
semiconductor heterostructures, while we are interested in the
statistical physics of clean systems in the thermodynamic limit.
Leijn\-se \cite{PhysRevB.87.125417} has more recently studied a square array
without degeneracy. This work did not address the possibility of
charge ordering and did not employ Monte Carlo simulations.
A two-dimensional (2D) layer with interactions and hopping was studied
within a Keldysh approach by Mitra \textit{et al}.\
\cite{PhysRevLett.97.236808}. Their interest was in ferromagnetic
order in the 2D layer and in the universality class of the non-equilibrium,
voltage-driven phase transition.

The structure of this paper is as follows: In Sec.\ \ref{sec.Method}, we
define the model and methods. Section \ref{sec.results.zero} gives a
comprehensive discussion of the phases and phase transitions at
zero temperature. Section \ref{sec.results.nonzero} presents
results for nonzero temperatures. We give a summary in
Sec.~\ref{sec:summary}.

\section{Model and methods}
\label{sec.Method}

\begin{figure}
	\includegraphics[width=0.65\columnwidth]{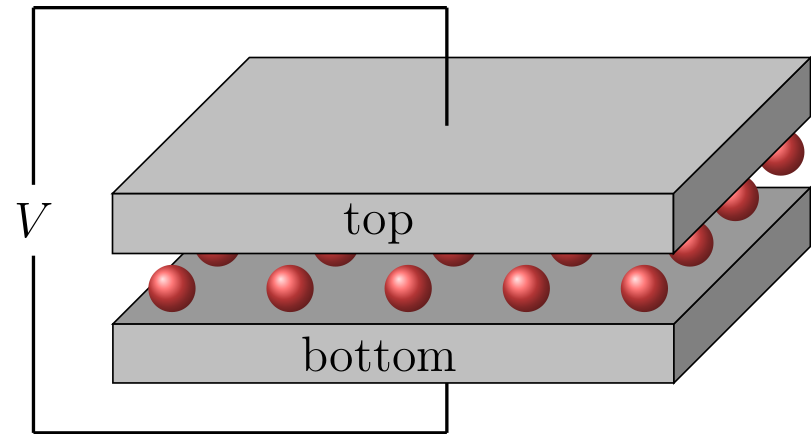}
	\caption{(Color online) Sketch of the model system, a square-lattice
	monolayer of quantum dots or molecules sandwiched between two conducting
	electrodes. A bias voltage $V$ is applied between the electrodes.}
	\label{fig:sketch_of_system}
\end{figure}

Our model consists of a 2D square lattice of quantum
dots or molecules sandwiched between two conducting electrodes, as sketched in
Fig.\ \ref{fig:sketch_of_system}. The Hamiltonian
$H = H_\text{leads} + H_\text{layer} + H_\text{tun}$
consists of the three terms
\begin{align}
H_\text{layer} &= E_d \sum_{i\sigma}{\hat n}_{i\sigma}
  + \frac{U_0}{2} \sum_i
  \sum_{\sigma\neq\sigma'} {\hat n}_{i\sigma} {\hat n}_{i\sigma'}
  \nonumber \\
& \quad {}+ U_1
  \sum_{\langle ij\rangle} \sum_{\sigma\sigma'}\,
  {\hat n}_{i\sigma} {\hat n}_{j\sigma'},
\label{eq:Hlayer} \\
H_\text{leads} &= \sum_{\alpha\mathbf{k}\sigma} \left(
  \varepsilon_\mathbf{k} - \mu_\alpha \right)
  a^\dagger_{\alpha\mathbf{k}\sigma} a_{\alpha\mathbf{k}\sigma},\\
H_\text{tun} &= \sum_{\mathbf{k} \sigma i \alpha} t_{\mathbf{k}i\alpha}\,
  a^\dagger_{\alpha\mathbf{k}\sigma} c_{i \sigma} +
  \text{H.c.},
\end{align}
where $E_d$ is the single-particle energy of the individual dots or molecules, $U_0$
is the on-site Coulomb interaction, $U_1$ is the nearest-neighbor Coulomb
interaction, $\varepsilon_\mathbf{k}$ is the dispersion of electrons in
the electrodes, $\mu_\alpha$ is the chemical potential in electrode
$\alpha=1,2$, and $t_{\mathbf{k}i\alpha}$
is the tunneling amplitude between the electrodes
and the monolayer. For simplicity, the tunneling amplitude
$\tilde t\equiv t_{\mathbf{k}i\alpha}$ and
the density of states of the electrodes are assumed to be constant.
$c_{i\sigma}$ is the electronic annihilation operator for a state in
the monolayer at site $i$ with quantum numbers $\sigma$, which could include the
spin but, importantly, may also include an orbital index, and
${\hat n}_{i\sigma}\equiv c^\dagger_{i\sigma} c_{i\sigma}$
is the corresponding number operator.
$a_{\alpha\mathbf{k}\sigma}$ is the annihilation operator for a state with
spin $\sigma$ and momentum $\mathbf{k}$ in electrode $\alpha$.
We use $U_1$ as our unit of energy and measure $E_d$
relative to the chemical potential in equilibrium.
We consider the limit $U_0 \to \infty$
so that each site can only be empty or singly occupied.
The voltage drop is assumed to be symmetric and the applied bias voltage
is given by $eV = \mu_1 - \mu_2$.

The Hamiltonian for the layer, Eq.\ (\ref{eq:Hlayer}), does not contain
intralayer hopping. For quantum dots, this situation is easily realized
by making the separation between dots sufficiently large. On the other hand,
molecular layers are typically closely packed. It is nevertheless possible
to reduce the overlap between the relevant orbitals of neighboring molecules
by choosing appropriate side groups. In the absence of wave-function overlap and
tunneling within the layer, the exchange interaction between different
sites also vanishes and we are left with the direct Coulomb interaction in Eq.\
(\ref{eq:Hlayer}). The spin thus only enters by causing a
twofold degeneracy. From a theoretical perspective, inclusion of
intralayer hopping would transform the system into an extended Hubbard model out
of equilibrium, a much more difficult problem. The methods we will discuss below
rely on the decomposition of the many-particle dynamics into
single-site processes (coupled by their dependence on the occupation of neighboring
sites). This would not be possible in the presence of intralayer hopping, which
would instead require us to consider the many-body eigenstates of an extended
Hubbard model.

The degeneracy of the occupied states, i.e., the number of possible
realizations of an occupied site, is denoted by $G$, while we assume the
unoccupied state to be non-degenerate. Hence, for $L^2$ lattice sites
($L$ is the linear size of the system), there are
$(G+1)^{L^2}$ possible many-particle states.
It is, however, advantageous to view the $G$ occupied states as a single
one and include the degeneracy factor $G$ explicitly in the equations.
The case of a single orbital per site, spin $1/2$, and
vanishing magnetic field corresponds to $G=2$.
A strong magnetic field that shifts one spin orientation up to high
energies would lead to $G=1$. Orbital degeneracies
and effective degeneracies due to vibrational modes \cite{Bolvin1995295,
*Bolvin1995355, J_Phys_I_France_5_6_747, PhysRevB.58.14238}
and local magnetic moments can result in larger values of $G$.
Degeneracies of both the occupied and the unoccupied states are easily included
and lead to the same results, except for overall constant factors,
where $G$ is now the ratio of the degeneracies of occupied and unoccupied
states. A degeneracy of what we call the unoccupied state is naturally
realized if the transition is not between an empty and a singly occupied
orbital but between a singly occupied and a doubly occupied orbital.
We conclude that it is meaningful to allow $G$ to take any positive real value.

\subsection{Master equation}

For weak tunneling to the electrodes but strong
interactions $U_0$ and $U_1$, the master-equation formalism is most
suitable \cite{bruus2004many, Breuer07, PhysRevB.77.195416, 2010Nanot..21A2001A}.
The master equation is the equation of motion for the reduced density operator
of the monolayer, $\rho_\text{red}=\Tr_\text{leads}\rho$, where
$\rho$ is the density operator of the whole system.
In the limit of weak tunneling, a perturbative
expansion in the tunneling amplitude $\tilde t$ can be employed. The 
sequential-tunneling approximation is obtained by truncating this expansion
after the second order. Furthermore, we make the standard assumption that the
monolayer and the electrodes are in a product state with the electrodes in
separate equilibrium at an early time $t_i\to -\infty$.
By suitably organizing the expansion (or, alternatively,
by a Markov assumption), we can make the resulting master equation
local in time \cite{Breuer07,PhysRevB.77.195416}.
The result is the Wangsness-Bloch-Redfield master equation~\cite{PhysRev.89.728, PhysRev.105.1206, Redfield},
\begin{align}
\frac{d\rho_\text{red}}{dt}
  =& -i\, \big[H_\text{layer},\rho_\text{red}(t)\big] \nonumber \\
& {}- \int_0^{\infty} d\tau\: \mathrm{Tr}_\text{leads}\,
  \Big[H_\text{tun},
  \Big[ e^{-i(H_\text{layer}+H_\text{leads})\tau} \nonumber \\
& {}\times H_\text{tun} e^{i(H_\text{layer}+H_\text{leads})\tau},
  \rho_\text{red}(t) \otimes \rho_\text{leads}^0 \Big]\Big] ,
\label{eq:WBR}
\end{align}
where $\rho_\text{leads}^0$ describes the initial equilibrium state of the 
electrodes and we have set $\hbar=1$. We focus on the stationary state, which
is obtained by setting $d\rho_\text{red}/dt=0$. The result is a linear algebraic
equation for $\rho_\text{red}$.

We employ the basis of occupation-number states in real space,
i.e., of eigenstates of all number operators ${\hat n}_{i\sigma}$.
The corresponding eigenvalues are denoted by $n_{i\sigma}=0,1$.
These states are
also eigenstates of $H_\text{layer}$. In this work, we assume that 
the stationary reduced density operator $\rho_\text{red}$ is diagonal in this
basis, i.e., we will neglect all coherences. In the following, we discuss the
conditions for this assumption to be valid.

First of all, coherences $\ketbra{a}{b}$ between states with
different total particle numbers $N_a$, $N_b$ dephase
nearly instantaneously due to superselection rules
\cite{PhysRev.88.101,*PhysRevD.1.3267,PhysRevD.26.1862,Giulini1995291}. Such
coherences are also seen to decouple from the diagonal components of 
$\rho_\text{red}$ (i.e., the probabilities) and from the coherences
between states with the same total particle number in Eq.\ (\ref{eq:WBR}).
Coherences of the latter type do couple to the diagonal components and are
generated with time even if they are not present in the initial state. 
The relevant processes involve the tunneling of an electron out of site
$i$ of the molecular layer into a typically adjacent site $i'$ in
electrode $\alpha$ and then back out of a site $j'$ in the electrode
into site $j$ in the molecular layer (or in the opposite temporal order).
Mathematically, they are controlled by 
the lesser and greater Green functions of lead electrons,
which appear in the integral term of the
master equation (\ref{eq:WBR}) \cite{PhysRevB.77.195416}.
In a clean system, the Green functions
for states close to the chemical potential decay in real space as
$\sin k_Fr'/k_Fr'$, where $r'$ is the distance between
points $i'$ and $j'$ in the electrode. 
This means that the terms generating non-local coherences
are small if the distance between neighboring molecules---or more precisely 
between the points in the electrodes connected to neighboring molecules by
tunneling---is large compared to the Fermi wavelength $\lambda_F=2\pi/k_F$
in the electrodes. Moreover, in the presence of
disorder in the leads, the Green functions are additionally cut off at the
scale of the mean free path $l$. We conclude that non-local coherences can be
neglected if the separation between neighboring molecules is large compared to
the Fermi wavelength or to the mean free path in the electrodes.

This leaves the possibility of coherences between states that only
differ by the local quantum numbers $\sigma$. Let us first consider the case
that $\sigma$ in Eq.\ (\ref{eq:Hlayer}) only refers to the
\textit{z}-component of the real spin. It is easy to check that
coherences between spin states decouple from the diagonal components
if the full Hamiltonian $H$ conserves
spin. This means that such coherences are not generated if they are not 
present in the initial state and, moreover, decay to zero if there is any
arbitrarily weak spin relaxation. This conclusion carries over to the case
with $\sigma$ containing additional (orbital) degrees of freedom: coherences
can be neglected if $H$ conserves the full set of quantum numbers
$\sigma$. This for example applies to a model involving molecules
with $p_x$ and $p_y$ (or $d_{xz}$ and $d_{yz}$) orbitals where the
interface is the \textit{xy} plane. Tunneling only occurs to lead orbitals with
the same mirror symmetries with regard to the \textit{xz} and \textit{yz}
planes so that the pseudo-spin distinguishing between the two orbitals is
conserved, in addition to the real spin.

If coherences are neglected, the master equation simplifies to rate equations
for the probabilities $P_a \equiv \erwart{a}{\rho_\text{red}}{a}$ of the
many-body states $\ket{a}$ of the monolayer:
\begin{equation}
	\frac{d}{dt} P_f = \sum_i \left( R_{i \to f} P_i - R_{f \to i} P_f
	\right).
	\label{eq:rate_equation}
\end{equation}
In the sequential-tunneling approximation, the rates take the
form \cite{PhysRevB.71.155403,PhysRevB.77.195416,PhysRevB.83.115416}
\begin{equation}
	R_{i \to f} = \sum_\alpha R^\alpha_{i \to f} ,
	\label{eq:seq_fullrates}
\end{equation}
with
\begin{align}
	R^\alpha_{i \to f} &= \frac{\tilde t^{\,2} D}{h} \sum_{j} \Big( G\,
	|c_{j}^\dagger|_{if}^2\, f_\alpha(E_d + z_j U_1) \nonumber \\
&\quad{}+ |c_{j}|_{if}^2\, \big[ 1-f_\alpha(-E_d -
	 z_j U_1) \big] \Big) \nonumber \\
&= \frac{\tilde t^{\,2} D}{h} \sum_{j} \big(G\,
	|c_{j}^\dagger|_{if}^2 + |c_{j}|_{if}^2 \big)\,
	f_\alpha(E_d + z_j U_1),
	\label{eq:seq_rates}
\end{align}
where $D$ is the density of states per electrode and spin,
$f_\alpha(x) \equiv f(x-\mu_\alpha)$ is the Fermi function $f(x)$, and
$|c_{j}^{\dagger}|_{if}^2 \equiv
| \langle  f | c_{j}^{\dagger} | i \rangle |^2$.
We have dropped the index $\sigma$ because  the rates do not depend on it and
the degeneracy $G$ is already included explicitly in the in-tunneling rates.
Furthermore, we have used that sequential tunneling only connects
many-body states $\ket{i}$, $\ket{f}$ that differ by a single electron 
at a single site $j$ and only depends on the local energy contributions
$E_d$ and $z_j U_1$, where $z_j$ is the number of occupied sites
neighboring $j$.

It is easy to check that in the case of $V=0$, i.e., for $\mu_1=\mu_2$, 
the rates $R_{i \to f}$ satisfy detailed balance so that the system relaxes 
into its equilibrium state at the temperature of the electrodes.
For $G=1$ and $V=0$, our system is equivalent to an Ising model
in equilibrium. The role of the Ising magnetic field is played by the
on-site energy $E_d$. The degeneracy $G$ can be absorbed into this
magnetic field as a temperature-dependent term, as we discuss below.

Our model satisfies a particle-hole symmetry. The symmetry
operation consists of interchanging in-tun\-ne\-ling and out-tun\-ne\-ling
processes and mapping the on-site energy
according to $E_d \to -4U_1-E_d$. For $G\neq 1$, the degeneracy of the
\emph{unoccupied} state becomes $G$ after the mapping. At the level of the rate
equations, this is equivalent to setting the degeneracy of the occupied state
to $1/G$ and multiplying $\tilde t^{\,2} D/h$ by $G$.

The staggered magnetization of an antiferromagnetic Ising model maps to
$\expect{n_A-n_B}$, where $n_A$ and $n_B$ are the occupation numbers 
of any site on the checkerboard sublattices $A$ and $B$, respectively.
The brackets $\expect{\cdots}$ denote the 
statistical average, over space and time, in the stationary state.
Due to $U_0\to\infty$, we have $0\le \expect{n_{A,B}} \le 1$.
We call $\expect{n_A-n_B}$ the checkerboard order parameter from now on.
The corresponding susceptibility $\chi$ is
\begin{equation}
\chi \equiv \big< ( n_A - n_B )^2 \big>
 - \left< n_A - n_B \right>^2 .
\end{equation}
Furthermore, we denote the total electron number in the molecular layer
by $N$, the number of nearest-neighbor bonds of type
$X \in \{ 00, 01, 10, 11\}$, corresponding to empty-empty, empty-occupied,
etc., by $N_X$, and the associated concentrations by $n=N/L^2$
and $n_X=N_X/L^2$. Lastly, the average current per site
from the monolayer into the electrode $\alpha$ is given by
\begin{equation}
  \expect{I^\alpha} = \frac{e}{L^2} \sum_{if} (N_f - N_i)\, R_{i \to f}^\alpha P_i,
  \label{eq:current}
\end{equation}
where $N_i$ ($N_f$) is the total electron number in the monolayer
in the initial (final) state.

\subsection{Mean-field approximation versus Monte Carlo simulations}
\label{sec.Method.MFME_and_MC}

We solve the rate equations (\ref{eq:rate_equation}) employing two
complementary methods. First, we apply a mean-field approximation at
the level of probabilities (mean-field master equation, MFME).
Specifically, we trace out all sites except for a single site
$j$ in the rate equations \eqref{eq:rate_equation}. The resulting
probability for site $j$ having the occupation number $n_j=0,1$ is
$P^j_{n_j} \equiv \sum_{\{n_i=0,1 | i \neq j\}} P_{\vec{n}}$,
where $\vec{n} = (n_1,\ldots,n_{L^2})$ represents a many-body
state of the whole layer in the occupation-number basis.
The rate equations then take the form
$d P^j_{n_j}/dt = F(P_{\vec{n}})$, where the right-hand side still depends on
the full configuration. The main approximation then consists of the
product ansatz $P_{\vec{n}} = \prod_j P^j_{n_j}$ in $F$.
This approximation leads to coupled equations for the single-site
probabilities $P^j_{n_j}$ for all sites $j$. Using $P^j_0+P^j_1=1$, these
are $L^2$ independent probabilities. To simplify the problem further,
we only consider specific spatial variations of the probabilities $P^j_{n_j}$.
Since we are interested in checkerboard order, we assume
$P^j_{n_j}$ to be the same for all $j$ on the same checkerboard sublattice
$A$ or $B$, i.e., $P^j_{n_j} = P^A_{n_A}$ for $j\in A$ and
$P^j_{n_j} = P^B_{n_B}$ for $j\in B$. This only permits
uniform and checkerboard-ordered solutions, where the uniform state
corresponds to $P^A_n=P^B_n$.
Since $P_0^s + P_1^s = 1$ for $s=A,B$, we have now reduced the
problem to finding two unknowns $P_1^A$ and $P_1^B$.
The product ansatz constitutes a mean-field-type approximation since
it replaces the spatial correlations included in $P_{\vec{n}}$ by much
simpler ones that only depend on the averaged occupation on each sublattice.
This approximation goes beyond a Hartree approximation, which would replace
the nearest-neighbor Coulomb interaction by the interaction with the average
charge density. Here, we retain the information that the sites are always
either occupied or unoccupied.

The resulting equation of motion for the probability of a site on
sublattice $s = A,B$, here denoted as site $0$, having the occupation $n_0$
reads as
\begin{align}
  \frac{d P_{n_0}^s}{dt} &= \sum_{n_1, \ldots, n_4} \big( R_{\ket{\bar{n}_0,z_0} \to   
  \ket{n_0,z_0}} P_{\bar{n}_0}^s \nonumber \\
    &\quad{}- R_{\ket{n_0,z_0} \to \ket{\bar{n}_0,z_0}} P_{n_0}^s \big)
    P_{n_1}^{\bar{s}} \cdots P_{n_4}^{\bar{s}}.
\label{eq:MFME}
\end{align}
In deriving this equation, we have used that under sequential
tunneling only the occupation number of a single site changes.
The transition rate for this process depends on whether an electron
tunnels in or out and on the number of occupied neighboring sites.
We can thus parametrize the rates $R_{\ket{n_0,z_0} \to \ket{\bar{n}_0,z_0}}$ by
the initial and final occupation numbers, $n_0$ and $\bar{n}_0\equiv 1-n_0$,
respectively, and the number $z_0 \equiv \sum_{i=1}^4 n_i$ of occupied neighboring
sites on the square lattice, enumerated by $i$.
In Eq.\ (\ref{eq:MFME}), $\bar{s} = B, A$ for sublattice $s = A, B$ since the
neighboring sites are always on the other sublattice.
Here and in the following, we use $\ket{n_j,z_j}$ as a short-hand notation for the
full many-body state $\ket{n_1,n_2,\ldots}$, which highlights the quantities that
affect the sequential tunneling at site $j$.
Stationary states are fixed points, which are obtained by setting
the time  derivatives to zero. We solve the equations numerically, discarding 
any unstable fixed points.
We note that this procedure is different from the one used in Ref.\
\cite{PhysRevB.87.125417}, which  is formulated in terms of the 
conditional occupation probabilities of the sites, depending on the 
occupation of their neighbors.

Second and foremost, we use Monte Carlo simulations.
While they are numerically more expensive,
they have the advantage of being free from approximations beyond those
made in the derivation of the sequential-tunneling rate equations
\eqref{eq:rate_equation}, apart from finite-size effects.
Moreover, the solutions are not restricted to uniform or checkerboard order.
As the system size is finite, we have to evaluate the average
$\expect{|n_A - n_B|}$ instead of $\left< n_A - n_B \right>$.
We consider linear system sizes $L$ between 16 and 16\,384
and periodic boundary conditions.

The straightforward algorithm for local updates is the following:
Randomly choose a site $j$. If this site is initially occupied
(unoccupied) the only possible transition is to the unoccupied
(occupied) state. Calculate the rate $R_j \equiv R_{i\to f}$
for this transition from Eqs.\ \eqref{eq:seq_fullrates}) and
\eqref{eq:seq_rates}; the result depends on
the occupations of the neighboring sites. Accept the
transition with the probability $R_{i\to f}/\max(R)$,
where $\max(R)= {2 G \tilde t^{\,2} D}/{h}$ is the
maximum possible rate.

This algorithm is highly inefficient if the rate $R_j$
chosen as described above often turns out to be small compared to
$\max(R)$ since then many Monte Carlo steps are rejected.
Hence, we instead employ the rejection-free update
scheme described in the following.
First, note that there are only $10$ distinct 
local updates in sequential tunneling, which
are enumerated by the initial occupation $n_j=0,1$ and the
number of occupied neighbors, $z_j=0,\ldots,4$.
We first randomly select one of the 10 update types with the proper branching
fraction $\tilde R_{n_j,z_j}/\sum_{n_k,z_k} \tilde R_{n_k,z_k}$, which is 
determined from the total rates of processes of each of the types,
$\tilde R_{n_j,z_j} \equiv R_{\ket{n_j,z_j} \to \ket{\bar{n}_j,z_j}}
\sum_p \delta_{n_p,n_j} \delta_{z_p,z_j}$,
where $p$ enumerates the sites of the system. These quantities are easily
updated in each Monte Carlo step. Then we randomly choose a site with occupation
$n_j$ and $z_j$ occupied neighbors. This step is rejection free since the
program keeps lists of all sites that are in each of the 10 states
$\ket{n_j,z_j}$.
Lastly, we update this site and advance the simulation time by 
the average waiting time $1 / \sum_{n_k,z_k} \tilde{R}_{n_k,z_k}$
\footnote{Kinetic simulations would require us to advance
the simulation time by an exponentially distributed time with
this mean. However, for the long-time limit
we are interested in, both prescriptions give the same results.}.
This algorithm is related to the one introduced by Bortz
\textit{et al}.\ \cite{BORTZ197510} and
independently by Gillespie \cite{GILLESPIE1976403, *doi:10.1021/j100540a008},
and further improved by Schulze \cite{PhysRevE.65.036704}.
We measure all times in units of $t_0 \equiv h/(\tilde t^{\,2} D)$, which
is the average waiting time between two tunneling events at a single site
if the transition is inside the bias window and $T=0$ and $G=1$.
To our knowledge, more efficient update algorithms, such as cluster updates
\cite{PhysRevLett.57.2607, PhysRevLett.62.361},
do not exist for the rates in Eq.\ \eqref{eq:seq_fullrates}, which do not
satisfy detailed balance.

\section{Results for zero temperature}
\label{sec.results.zero}

\begin{figure}
	\includegraphics[width=\columnwidth,keepaspectratio=true]{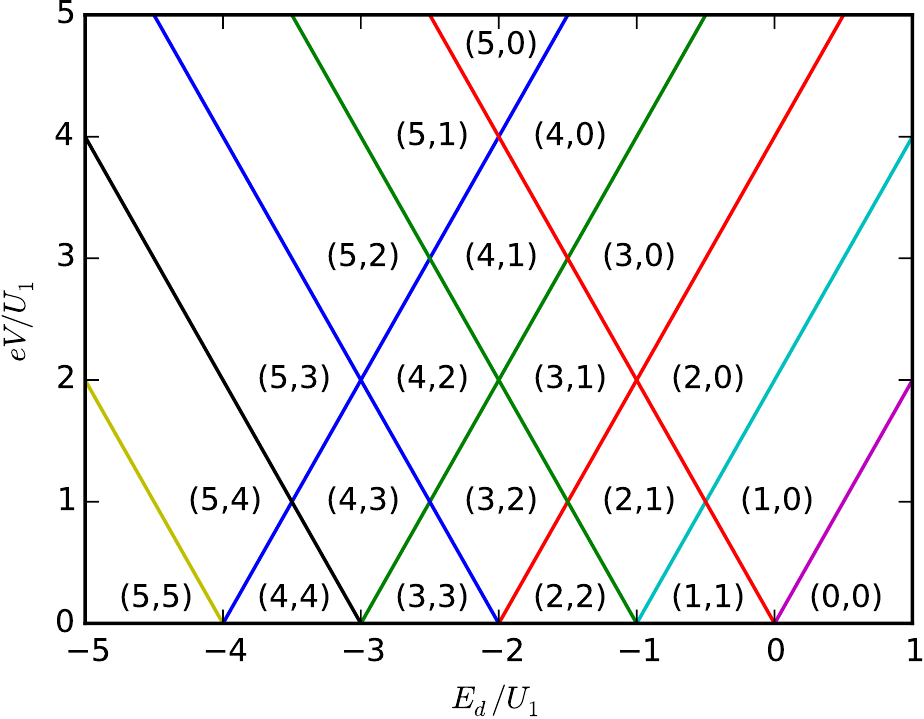}
	\caption{(Color online)	Crossings of the transition energies of the
	square-lattice monolayer and the chemical potentials
	of the electrodes, as functions of the
	on-site energy $E_d$ and the bias voltage $V$. In the limit $T \to
	0$, the transition rates $R_{i\to f}$ are constant within each
	region. The labels $(m,n)$ specify the numbers of
	transition energies below the chemical potentials,
	$m$ for the top electrode 1 and $n$ for the bottom electrode~2.}
	\label{fig:phase_diagramm_T0_energy_crossings}
\end{figure}

\begin{figure}
	\includegraphics[width=0.85\columnwidth]{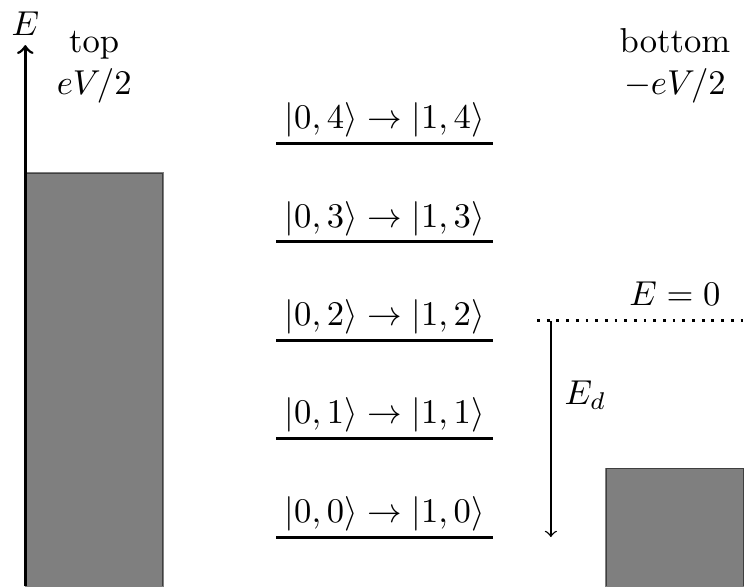}
	\caption{Sketch of the chemical potentials and the transition energies
	for in-tunneling processes for the region $(4,1)$ of
	Fig.\ \ref{fig:phase_diagramm_T0_energy_crossings}.
	$\ket{n_j,z_j}$ denotes a many-particle state for which the	
	site $j$ involved in the tunneling has occupation number $n_j$ and $z_j$
	occupied neighbors (all other occupation numbers are irrelevant for
	this process and are suppressed).}
	\label{fig:energy_bias_diagram}
\end{figure}

\begin{figure}
	\includegraphics[width=\columnwidth,keepaspectratio=true]{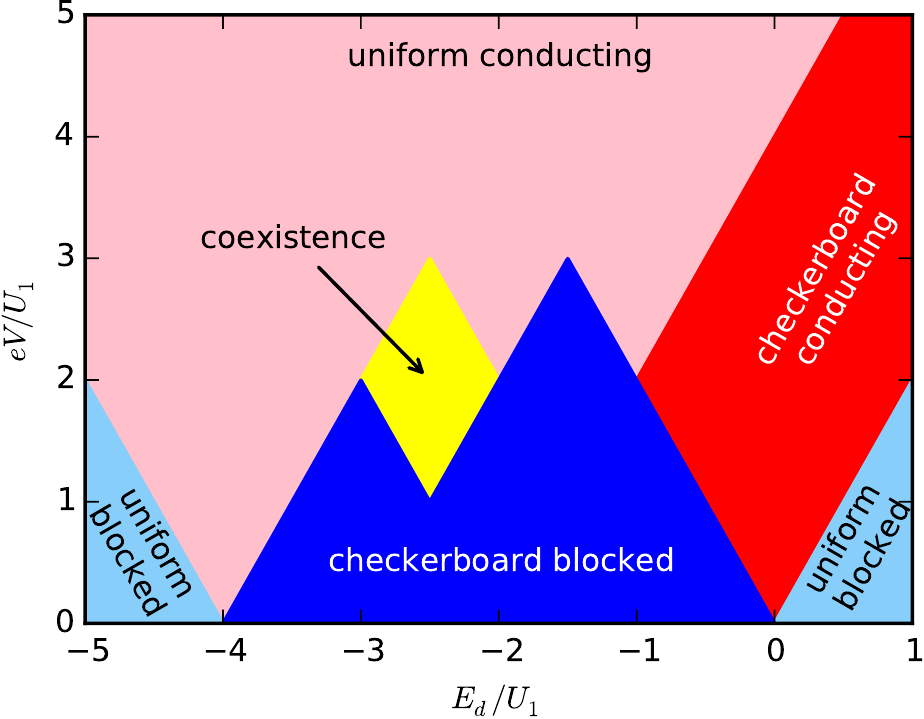}
	\caption{(Color online) Phase diagram obtained from the MFME for a
	monolayer with degeneracy $G=2$ and temperature $T=0$. The phases for
	negative bias voltage, $eV/U_1<0$, are the mirror image of the phases
	shown. The terms ``uniform'' and
	``checkerboard'' refer to the checkerboard order parameter
	$\expect{ n_A-n_B }$, where $n_A$ and $n_B$ are the
	occupation numbers per site on sublattices $A$ and $B$,
	respectively. This order parameter is zero (nonzero) in the uniform
	(checkerboard) phases. The term ``conducting'' (``blocked'')
	characterizes phases that carry (do not carry) a current through the
	monolayer. In the region labeled ``coexistence,'' checkerboard
	blocked and uniform conducting stationary states coexist.}
	\label{fig:phase_diagramm_T0_Cluster_MF_ME}
\end{figure}

In the limit of zero temperature,
the Fermi distribution becomes a step function and thus
the rates change discontinuously. Consequently, the stationary
state is the same for all values of
$E_d$ and $eV$ within each of the regions defined in Fig.\
\ref{fig:phase_diagramm_T0_energy_crossings}. This allows us to give 
a complete discussion of all possible stationary states.
The regions are labeled by the numbers of transition energies
below the chemical potentials of the two
electrodes. As an example, Fig.\ \ref{fig:energy_bias_diagram} shows the
relevant energies for the region $(4,1)$: four transitions lie below the
chemical potential of the top electrode but only one transition lies
below the chemical potential of the bottom electrode.

To get an overview, we present in Fig.\
\ref{fig:phase_diagramm_T0_Cluster_MF_ME} the MFME phase diagram for
the case $G=2$.
We find all possible combinations of phases with and without
checkerboard charge order and with and without a charge current
through the monolayer. In the presence of checkerboard order,
the MFME has two stationary solutions,  which
are related by interchanging the two sublattices.
In the region labeled ``coexistence,'' the
stationary MFME has both checkerboard blocked and uniform conducting
solutions. In the following, we discuss each phase based on
rigorous results and Monte Carlo simulations.

\subsection{Uniform conducting phase at large bias}

For fixed on-site energy $E_d$ and sufficiently high bias voltage
$V$, the system is always in the region $(5,0)$ (see Fig.\
\ref{fig:phase_diagramm_T0_energy_crossings}). In this regime, an analytical
solution of the rate equations (\ref{eq:rate_equation})
is possible. Since the in-tunneling and out-tunneling rates are
independent of the occupation numbers of the neighboring sites, the dynamics
of the individual sites is decoupled. In the stationary state,
the probabilities of a site $j$ being occupied or unoccupied are
$P(n_j=0) = 1/(G+1)$ and $P(n_j=1) = G/(G+1)$, respectively.
These are the same values one finds for the \emph{equilibrium} 
($V=0$) state in the limit $T\to\infty$.
The average current per site is
$\expect{I^1} = e D \tilde t^{\,2} G/h(G+1)$.
The system is clearly in the uniform conducting phase.
The solution of the rate equations \eqref{eq:rate_equation}
is given by the product of the aforementioned single-site
probabilities since the sites are decoupled.

\subsection{Quasi-equilibrium, blocked phases}
\label{sec.results.zero.QEB}

All regions connected to the zero-bias line, i.e., the regions
$(m,m)$ and $(m+1,m)$
in Fig.\ \ref{fig:phase_diagramm_T0_energy_crossings}, can be mapped
onto an equilibrium Ising-type model with degeneracy $G$ of one of the
states and an applied magnetic field, as we show in the following.
The equilibrium Ising model with $G=1$ has of course been
investigated thoroughly \cite{pathria2011statistical_mab216023733,
doi.10.1007.BF01325537, PhysRevB.16.1168}.
The mapping between our monolayer Hamiltonian and the Ising
Hamiltonian
\begin{equation}
	H_\text{Ising} = -J \sum_{\left< ij \right>} S_i S_j - B \sum_i S_i
\end{equation}
reads as
\begin{align}
	S_i &= 2n_i-1 = \pm 1,\\
	J &= -U_1/4, \\
	B &= -E_d/2-U_1 .
	\label{eq:Ising_B}
\end{align}
The state $S_i=+1$ has degeneracy $G$.

We first consider the regions $(m,m)$, which contain parts of the line
$eV/U_1=0$. Recall that the $T=0$ stationary state is the same
throughout each region. It is thus sufficient to investigate the case
$eV/U_1=0$, which corresponds to the model in equilibrium, in the limit $T\to
0$. But this state is just the ground state of $H_\text{Ising}$. For $E_d/U_1<-4$,
this is the fully occupied state. For $-4 < E_d/U_1 < 0$, there are two
degenerate ground states with
checkerboard charge order with one sublattice occupied and the other
unoccupied. Finally, for $E_d/U_1>0$ the ground state is completely empty.
The degeneracy $G$ is irrelevant in these cases since the energy of the
microstates
does not depend on it. The corresponding currents [Eq.\ \eqref{eq:current}]
vanish in all these regions since all transition rates out of the respective
ground states go to zero for $T\to 0$ \footnote{One might ask
whether the rate equations really describe relaxation into these
ground states. This is the case if we
set $V=0$ and then take the limit $T\to 0$ since the system is ergodic for any
$T>0$}.

Next, we turn to the regions $(m+1,m)$, which touch the line $eV/U_1=0$ at
a single point. While the stationary state is still the same throughout each
of these regions, the equilibrium state now lies at a corner of the region,
which may be distinct from its interior. These corner points at $V=0$ have
fine-tuned values of $E_d/U_1=-4,-3,-2,-1,0$ (see Fig.\
\ref{fig:phase_diagramm_T0_energy_crossings}), and correspond to one transition
energy being resonant with the chemical potential, which is the same for
both electrodes.
One can see from Eq.\ \eqref{eq:seq_rates} that the rates in the interior,
i.e., for $V\neq 0$, and at the corner,
i.e., for $V=0$, have the same limit for $T\to 0$. This
result relies on the symmetric coupling to the electrodes: for $V=0$, the
Fermi functions involving the resonant transition approach $1/2$ for both
electrodes in the limit $T\to 0$, whereas for $V\neq 0$, one of them approaches
unity and the other zero. Since for symmetric coupling only their sum enters,
the results are the same.
For the regions $(2,1)$, $(3,2)$, and $(4,3)$, the stationary states
are thus identical to the ground states for fine-tuned on-site energies
$E_d/U_1=-1$, $-2$, and $-3$, respectively.
However, these ground states are not
different from the rest of the range $-4<E_d/U_1<0$: they are the two
states with checkerboard order. The current again
vanishes, by the same argument as above.
The regions $(1,0)$ and $(5,4)$ are special in that their corner
points at $V=0$ lie right on the transition between different ground
states. We will investigate these cases in Sec.~\ref{sec.results.zero.CCG}.

In summary, in the regions $(1,1)$, $(2,1)$, $(2,2)$, $(3,2)$, $(3,3)$,
$(4,3)$,  and $(4,4)$ we find checkerboard charge order and 
vanishing current, in agreement with the MFME phase diagram in Fig.\
\ref{fig:phase_diagramm_T0_Cluster_MF_ME}. This phase can be understood
in terms of
Coulomb blockade due to the nearest-neighbor repulsion $U_1$.
In the region $(5,5)$, the sites are fully occupied and the current vanishes.
This is the Coulomb-blockade regime due to the on-site repulsion
$U_0$. Finally, in the region $(0,0)$ we find an empty lattice and
vanishing current.

\subsection{Degeneracy-driven phase transitions in the conducting phases}
\label{sec.results.zero.CCG}

We now consider the regions $(1,0)$, $(2,0)$, $(3,0)$, and $(4,0)$ and their
particle-hole-symmetry partners $(5,4)$, $(5,3)$, $(5,2)$, and $(5,1)$ in Fig.\
\ref{fig:phase_diagramm_T0_energy_crossings}.
The regions $(1,0)$ and $(5,4)$ are interesting since the
discussion in Sec.\ \ref{sec.results.zero.QEB} suggests that their stationary
states inherit properties from an equilibrium
Ising model fine tuned to its critical
point. Moreover, for the region $(1,0)$ [as well as $(2,0)$]
the MFME for $G=2$ predicts a state with
checkerboard charge order that is nevertheless conducting (see Fig.\
\ref{fig:phase_diagramm_T0_Cluster_MF_ME}). On the other hand, the state in
region $(5,4)$ is uniform and conducting, according to the MFME.
Since the region $(5,4)$ with degeneracy $G$ is equivalent to the region
$(1,0)$ with degeneracy $1/G$ by a particle-hole transformation, 
the MFME results imply the existence of a phase transition between
uniform and checkerboard states as a function of $G$.
Indeed, by determining the stationary state of the MFME for various $G$,
we find the critical degeneracy $G_c \approx 1.054$. It is then
interesting to characterize this phase transition without making a
mean-field approximation. In particular, we want to determine its universality
class.

Before turning to the simulations, we first present an analytical
estimate for the critical degeneracy $G_c$ for the region $(1,0)$. The
critical value for the region $(5,4)$ is then just $1/G_c$. As noted in Sec.\
\ref{sec.results.zero.QEB}, in the limit $T\to 0$ the rates in the interior of
region $(1,0)$ are the same as at the corner point $eV/U_1=0$, $E_d/U_1=0$.
The critical value $G_c$ can be related to the critical magnetic field
$B_c$ of an antiferromagnetic Ising model. We set $k_B=1$ and
consider the partition function $Z=\sum_a G^{N_a} e^{-E_a/T}$, where
$E_a$ is the energy of microstate $\ket{a}$ and
$N_a$ is the number of occupied sites in this microstate.
The energy is given by $E_a = N_a E_d + Z_a U_1$, where $Z_a$ is the number
$N_{11}$ of bonds between two occupied neighboring sites in
microstate $\ket{a}$. The partition function can be written as
\begin{equation}
	Z = \sum_a \exp \left( -\frac{ N_a\,(E_d - T\ln G) 
	  + Z_a U_1}{T} \right).
	\label{eq:CCG:partition_function}
\end{equation}
According to Eq.\ \eqref{eq:Ising_B}, the Ising magnetic field is
now $B = -(E_d-T\ln G)/2 - U_1$. At the corner point
of the region $(1,0)$, we have $E_d=0$. The equivalent
Ising model shows a phase transition between checkerboard and uniform
order as a function of magnetic field. The critical field $|B|=B_c$ is
determined by the temperature $T$ and the coupling constant $J=-U_1/4$. Taking
into account that $B$ is negative for small $T$, we can then write the critical
degeneracy as
\begin{equation}
	G_c(E_d,U_1,T) = \exp\left(\frac{-2B_c(U_1,T)+2U_1}{T}\right) .
\end{equation}
This expression is exact but, to the best of our knowledge, the analytical form
of the function $B_c$ is not known \cite{doi.10.1007.BF01325537, PhysRevB.37.1766}.
The conjectured low-temperature expansion
proposed by M\"uller-Hartmann and Zittartz \cite{doi.10.1007.BF01325537},
$B_c \cong 4|J| - T\ln 2 = U_1 - T\ln 2$, gives $\lim_{T\to 0} G_c = e^{2\ln 2}
= 4$.

\begin{figure}
\includegraphics[width=\columnwidth,keepaspectratio=true]{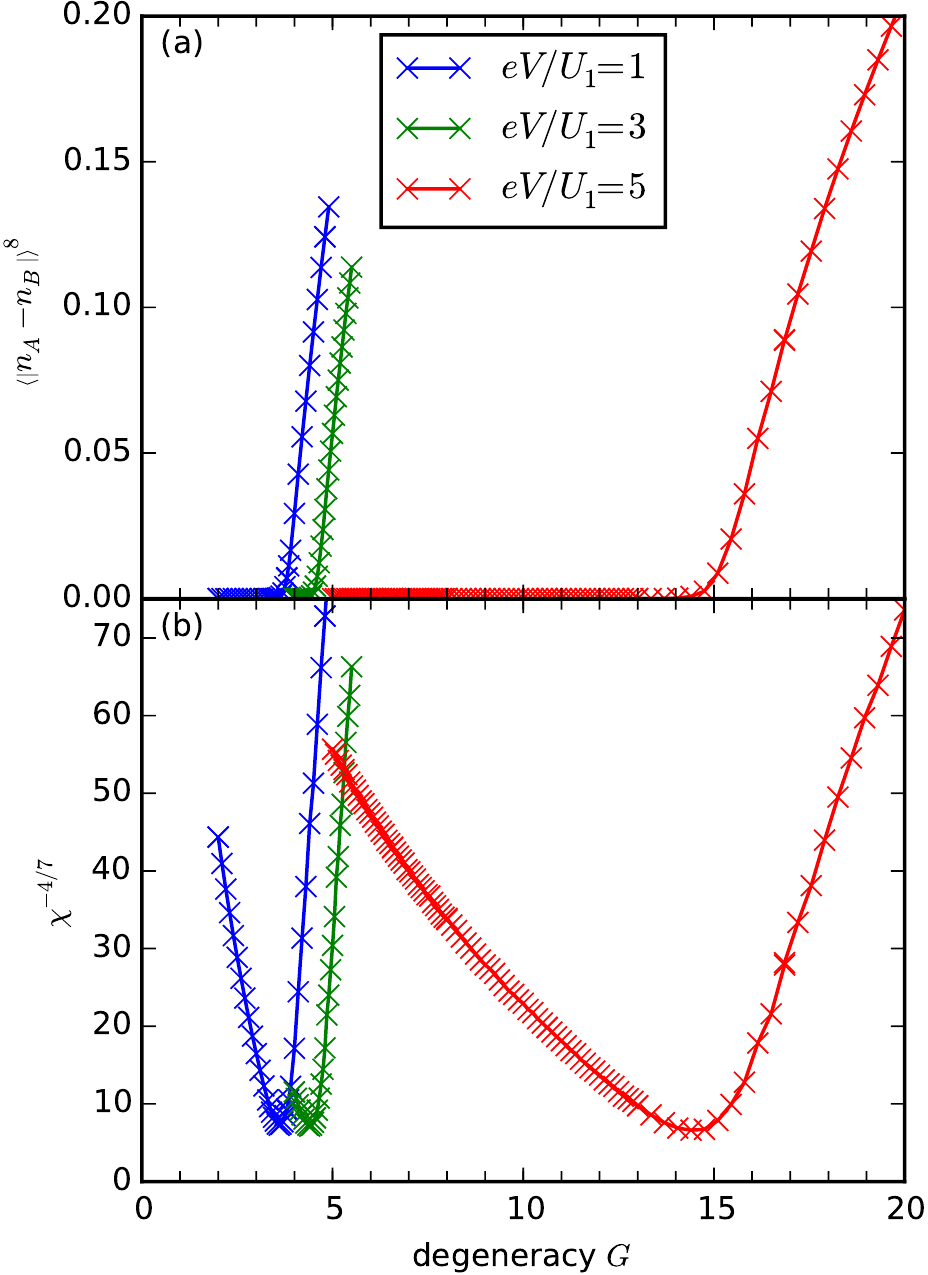}
	\caption{(Color online) Monte Carlo results for (a) the checkerboard
	order parameter $\expect{ \left| n_A-n_B \right| }$ to the power $8$ and
	(b) the corresponding susceptibility to the power $-4/7$
	as functions of the degeneracy $G$ of the occupied
	states. The values $eV/U_1=1,3,5$ correspond to
	the regions $(1,0)$, $(2,0)$, and $(3,0)$ in Fig.\
	\ref{fig:phase_diagramm_T0_energy_crossings}, respectively.
	The remaining parameters are $T/U_1 = E_d/U_1 = 0$ and $L=64$.
	The results are consistent with the 2D Ising universality class.}
	\label{fig:CCG_T0_Chi47_op8}
\end{figure}

The above discussion is based on the mapping to an equilibrium Ising
model. This is not possible for the regions $(2,0)$ and $(3,0)$, which should,
however, be physically similar to $(1,0)$ since all out-tunneling
transitions are energetically possible while only some in-tunneling transitions
are allowed. [The region $(4,0)$ will be discussed below.]
We have performed Monte Carlo simulations to study the degeneracy-driven
transition in these regions. Figure \ref{fig:CCG_T0_Chi47_op8} shows
the checkerboard order parameter raised to the power $8$ and the
corresponding susceptibility to the power $-4/7$. For all three regions
$(1,0)$, $(2,0)$, and $(3,0)$, we find a transition to checkerboard order for
increasing degeneracy $G$.
The critical degeneracy $G_c\approx 3.6$ for the region $(1,0)$  is slightly 
smaller than the value of $G_c\approx 4$ based on the
low-temperature expansion of Ref.\ \cite{doi.10.1007.BF01325537}.
On the other hand, the MFME prediction, $G_c\approx 1.054$,
is clearly much too small. Indeed, for the case of $G=2$ with pure
spin degeneracy, our simulations do not find checkerboard order in
any conducting region, in contrast to the MFME phase diagram in
Fig.~\ref{fig:phase_diagramm_T0_Cluster_MF_ME}.

Figure \ref{fig:CCG_T0_Chi47_op8}
shows some finite-size effects, in particular in the
susceptibility. Nevertheless, the transitions in the regions $(1,0)$,
$(2,0)$, and $(3,0)$ are consistent
with critical exponents of $1/8$ for the order parameter and $-4/7$
for the susceptibility, respectively, and thus with 2D Ising critical
behavior \cite{pathria2011statistical_mab216023733}.
The non-equilibrium transition is indeed expected to belong to the
classical 2D Ising universality class based on the arguments of 
Ref.\ \cite{PhysRevLett.97.236808}: integrating out the microscopic
degrees of freedom, one obtains
a description in terms of a classical field coupled to noise. Generically,
the spectrum of this noise is nonzero in the zero-frequency limit, which
corresponds to the Ising universality class (``model A'' in the terminology of
Hohenberg and Halperin \cite{RevModPhys.49.435}). Newer works show that this
is indeed true for a scalar (Ising) model, such as ours,
but not for multi-component order
parameters~\cite{PhysRevLett.110.195301, *PhysRevB.89.134310}.

Moreover, $G_c$ increases with increasing bias voltage $V$, i.e., from region
$(1,0)$ to $(2,0)$ and even more to $(3,0)$. This dependence can be understood
as follows: for a perfect
checkerboard-ordered state, a current flows through the occupied
sublattice, as the out-tunneling transition is in the bias window.
In contrast, the in-tunneling transition needed to fill a site on the
empty sublattice, i.e., to create an occupied defect, is forbidden,
except when there are enough empty defects at the surrounding
sites on the occupied sublattice. Raising the bias voltage lowers the
required number of empty defects and thus makes it easier to destroy
the checkerboard order. On the other hand, raising $G$ increases the
in-tunneling rate for filling an empty site, i.e., for removing
an empty defect, which stabilizes the
order~\footnote{Raising $G$ also increases the rate for creating an
occupied defect on the empty sublattice, which destabilizes the order, but
for this to happen more than one empty defect at neighboring sites is
needed, which makes the overall creation rate of such defects much 
smaller than the overall rate for removing empty defects.}.

In the region $(4,0)$, checkerboard order is even more strongly
destabilized than in the previous cases since an
occupied defect on the empty sublattice is possible as soon as
one empty defect on the occupied sublattice exists. Furthermore, 
the average concentration of empty defects on the occupied
sublattice is high since the out-tunneling transition is
in the bias window.
This concentration is suppressed by a large $G$ but at the same
time the creation rate for occupied defects neighboring an empty defect
is enhanced. The latter effect evidently prevents charge
order even at large $G$. Our simulations do not show any sign
of a transition for $G$ up to $10^8$.

The regions $(5,4)$, $(5,3)$, $(5,2)$, and $(5,1)$ are related to the
regions $(1,0)$, $(2,0)$, $(3,0)$, and $(4,0)$, respectively, by interchanging
empty and occupied states and replacing $G$ by $1/G$. This means that
degeneracy-driven transitions in regions $(5,4)$, $(5,3)$, and $(5,2)$ take
place at critical values $G_c<1$, which correspond to higher degeneracy of the
empty state compared to the occupied state of a single site (see the
discussion of $G$ in Sec.~\ref{sec.Method}).

\subsection{Absorbing phase transitions}
\label{sec.results.zero.absorb}

We now turn to the three diamond-shaped regions highest in the bias voltage,
i.e., regions $(3,1)$, its particle-hole-symmetry partner
$(4,2)$, and $(4,1)$ in Fig.\ \ref{fig:phase_diagramm_T0_energy_crossings}.
In all three and indeed in all regions $(m,n)$ with $m<5$ and $n>0$,
the rates for the transitions $\ket{1,0} \to \ket{0,0}$ and
$\ket{0,4} \to \ket{1,4}$ both vanish (for the notation $\ket{n_j,z_j}$,
see Fig.\ \ref{fig:energy_bias_diagram}). Consequently, there
are no allowed transitions out of the perfect checkerboard states.
This is the defining property of \emph{absorbing} states
\cite{2000AdPhy..49..815H,RevModPhys.76.663}. Absorbing states are necessarily
stationary, but for an infinite system
it is possible that the stationary state approached from nearly all
(namely, all except for a fraction that vanishes in the thermodynamic
limit) initial states is not one of the absorbing states.
Such a non-absorbing stationary state is called \emph{active}.
If an active state only exists in part of the parameter range, then
an absorbing-to-active phase transition has to
occur~\cite{2000AdPhy..49..815H,RevModPhys.76.663}.

Our Monte Carlo simulations suggest that the stationary state in regions
$(3,1)$ and $(4,2)$ is an absorbing checkerboard state for $G=1$ but
is a uniform, and thus active, state in the region $(4,1)$. However,
the results have to be analyzed with care since for any finite system
the simulation will eventually end up in one of the absorbing states,
possibly after a very long time.

To check that the regions really lie on different sides of an absorbing
phase transition, it is desirably to tune continuously through the purported
transition. In addition, this would allow us to determine its universality
class. However, the naive idea of fixing the temperature $T$ to a
small nonzero value and tuning the bias voltage $V$ does not work since at
$T>0$ the transitions out of the perfect checkerboard states 
occur with nonzero rates so that these states are no longer absorbing.
Instead, we define $\Delta V \equiv V-V_c$, where $V_c$ is a bias voltage on
the boundary between the regions $(4,1)$ and $(3,1)$
for a suitably chosen $E_d$. The boundary between regions $(4,1)$ and
$(4,2)$ is analogous. The rates $R_{\ket{0,4} \to \ket{1,4}}$ and $R_{\ket{1,0}
\to \ket{0,0}}$, which make the checkerboard state non-absorbing,
are then proportional to the Fermi function $f(U_1 + \Delta V)$.
These rates are tuned to zero by letting $T \to 0$.
On the other hand, the rates $R_{\ket{0,1} \to \ket{1,1}}$ and
$R_{\ket{1,1} \to \ket{0,1}}$ are proportional to $f(\Delta V)$ and
are kept constant by taking $\Delta V\to 0$ while keeping $\Delta V/T$ fixed.

Now, being able to tune our system continuously
by changing $\Delta V/T$, we use a dynamical-scaling
analysis \cite{PhysRevLett.18.891, *FERRELL1968565, PhysRev.177.952,
GRASSBERGER1979373, PhysRevE.56.5101, 2000AdPhy..49..815H}
to clarify the occurring phases and transitions.
Two standard critical exponents of an absorbing phase transition are
defined by the scaling relations~\cite{PhysRevE.56.5101, 2000AdPhy..49..815H}
\begin{align}
P_\text{surv} &\sim t^{-\delta},
\label{eq:Pstdelta.2} \\
\rho_\text{act} \equiv n_{00} + n_{11} &\sim t^\Theta
\end{align}
for large times $t$.
They pertain to a system prepared in an initial state 
that differs from an absorbing (checkerboard)
states by a localized defect. $P_\text{surv}$ is the 
survival probability, i.e., the probability that the system
does not reach an absorbing state until the time $t$, and
$\rho_\text{act}=n_{00}+n_{11}$
is the density of active sites \cite{2000AdPhy..49..815H},
which in our model correspond to
empty-empty and occupied-occupied nearest-neighbor bonds. Note that
both exponents are free from finite-size effects
since in our simulations the lattice was always
larger than any grown cluster of active sites.

\begin{figure}
\includegraphics[width=\columnwidth,keepaspectratio=true]{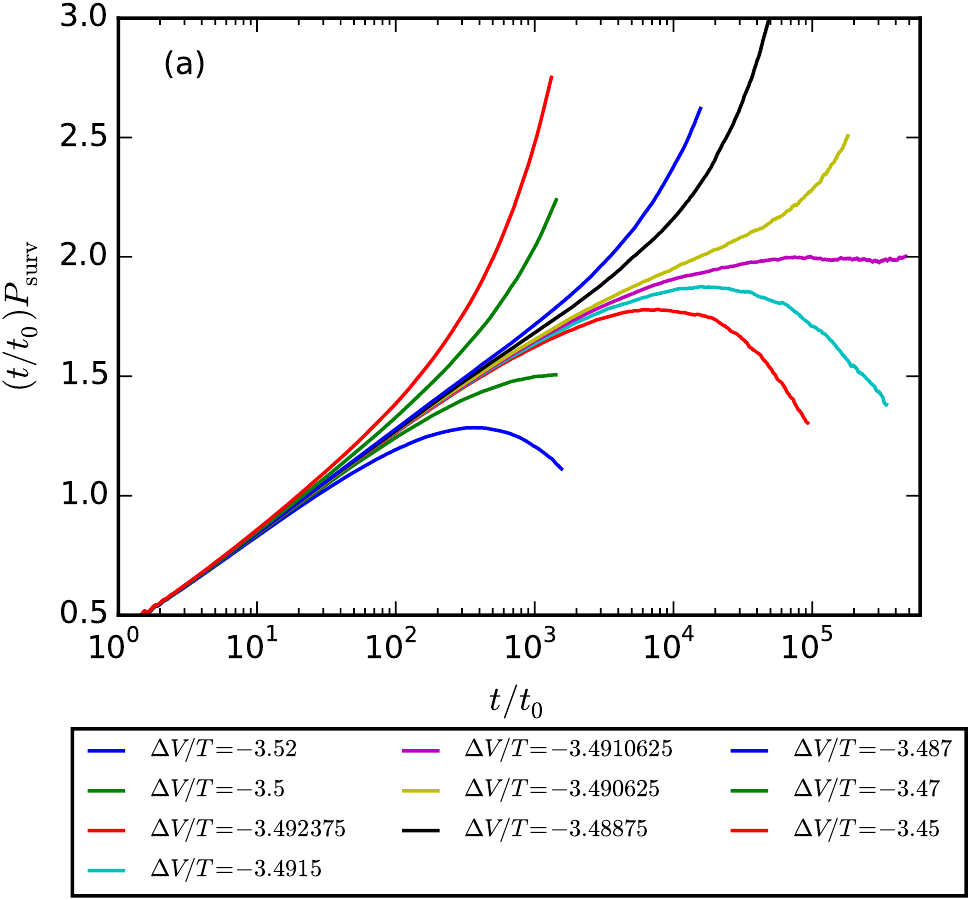}\\[2ex]
\includegraphics[width=\columnwidth,keepaspectratio=true]{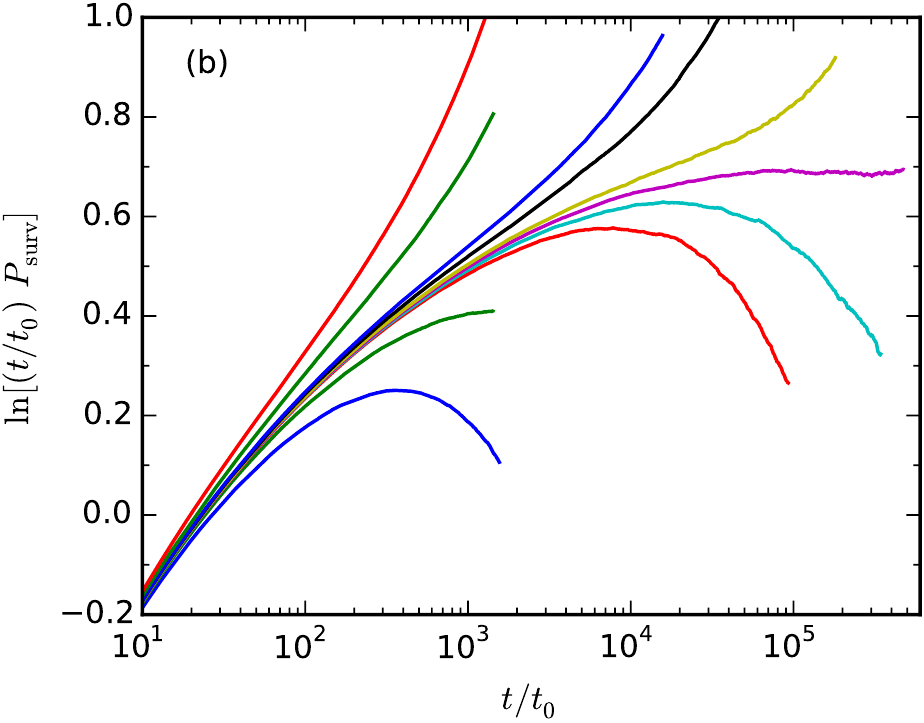}
	\caption{(Color online) Survival probability $P_\mathrm{surv}$ multiplied
	by $t/t_0$ vs.\ time $t/t_0$ in (a) single- and (b)
	double-logarithmic plots.
	The results have been obtained from Monte Carlo simulations for
	$G=1$ and various values of $\Delta V/T$ (see text),	
	starting with a single site deviating from the checkerboard state.
	The unit of time is $t_0 \equiv h/(\tilde t^{\,2} D)$.}
	\label{fig:survival_single_start}
\end{figure}

\begin{figure}
\includegraphics[width=\columnwidth,keepaspectratio=true]{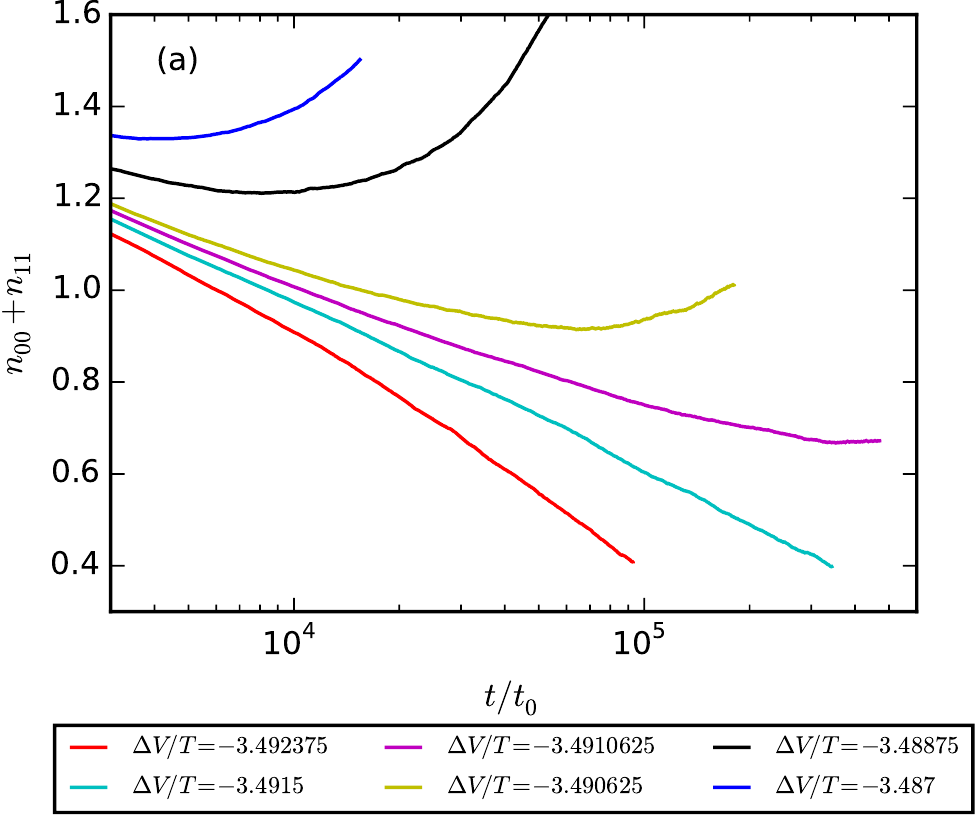}\\[2ex]
\includegraphics[width=\columnwidth,keepaspectratio=true]{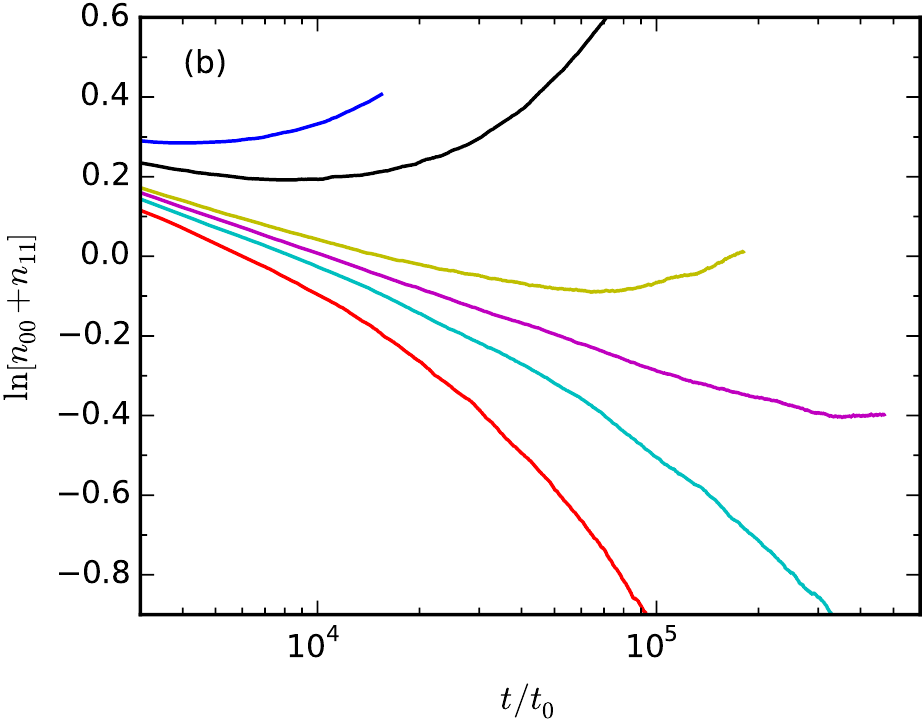}
	\caption{(Color online) Concentration of active bonds vs.\ time $t/t_0$
	in (a) single- and (b) double-logarithmic plots.
	The results have been obtained from Monte Carlo simulations for
	$G=1$ and various values of $\Delta V/T$ (see text),	
	starting with a single site deviating from the checkerboard state.}
	\label{fig:active_single_start}
\end{figure}

We concentrate on simulations for $G=1$.
Figure \ref{fig:survival_single_start} shows results for $P_\mathrm{surv}$.
Since the exponent $\delta$ is close to unity, we have plotted
$P_\mathrm{surv}$ multiplied by $t/t_0$.
Figure \ref{fig:active_single_start} shows results for
$\rho_\text{act}=n_{00}+n_{11}$.
We find a clear transition at $\Delta V/T \approx -3.49$ that agrees with
the 2D $\mathbb{Z}_2$ symmetric directed-percolation universality class
(DP2)~\cite{2000AdPhy..49..815H,PhysRevE.83.011114, PhysRevE.55.219}.
Dornic \textit{et al}.~\cite{PhysRevLett.87.045701} conjecture
that there is a mapping between the DP2 universality class and
a generalized voter model.
The latter has an upper critical dimension of two so that one expects
mean-field exponents $\delta=1$ and $\Theta=0$ with logarithmic
corrections \cite{PhysRevE.83.011114, PhysRevE.55.219}.
Previous work supports either mean-field behavior with logarithmic
corrections or exponents rather close to the mean-field ones
\cite{PhysRevE.83.011114, PhysRevE.55.219}, where the best estimate
is $\delta = 0.900(15)$ and $\Theta = -0.100(25)$~\cite{PhysRevE.83.011114}.

While our focus is not on this debate, we briefly comment on the
critical behavior. If $P_\mathrm{surv}$ followed
mean-field scaling with logarithmic corrections,
Fig.\ \ref{fig:survival_single_start}(a) would show a straight line for large
$t$, at the critical value of $\Delta V/T$. If $P_\mathrm{surv}$ instead satisfied
the power law Eq.\ (\ref{eq:Pstdelta.2}),
Fig.\ \ref{fig:survival_single_start}(b) would show a straight line.
Similarly, Fig.\ \ref{fig:active_single_start}(a)
[Fig.\ \ref{fig:active_single_start}(b)] would show a straight line at the critical
$\Delta V/T$ if $n_{00}+n_{11}$ showed mean-field scaling with logarithmic
corrections (power-law scaling).
While $t P_\mathrm{surv}$ in Fig.\ \ref{fig:survival_single_start} appears to
depend logarithmically on $t$ for small $t$ [note the straight line in Fig.\
\ref{fig:active_single_start}(a)],
$P_\mathrm{surv}$ crosses over to pure mean-field scaling with exponent $\delta=1$
and without logarithmic corrections
for $t/t_0 \gtrsim 10^4$ (horizontal line in both panels for
$\Delta V/T = -3.4910625$).
On the other hand, the scaling of $n_{00}+n_{11}$ shown in
Fig.\ \ref{fig:active_single_start} does not clearly discriminate between the
two scaling forms. The data for $\Delta V/T = -3.4910625$ are not consistent
with the mean-field exponent $\Theta=0$ without logarithmic corrections, i.e.,
with $n_{00}+n_{11}=\mathrm{const}$, for $t/t_0>10^5$, unlike in
Fig.\ \ref{fig:survival_single_start}.
We suggest that either the large-$t$ scaling regime has not been reached in
our simulations or the corrections are more complicated than a simple logarithm
$\ln(t/t_0)$.




\begin{figure}
\includegraphics[width=\columnwidth,keepaspectratio=true]{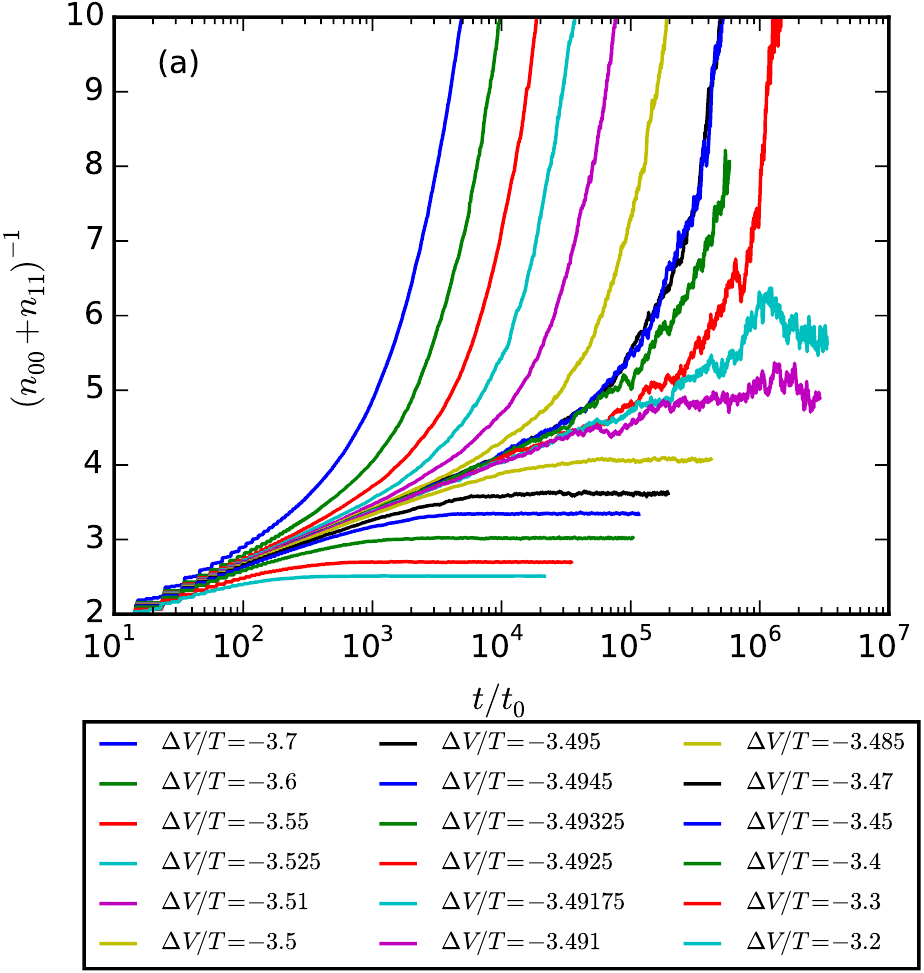}\\[2ex]
\includegraphics[width=\columnwidth,keepaspectratio=true]{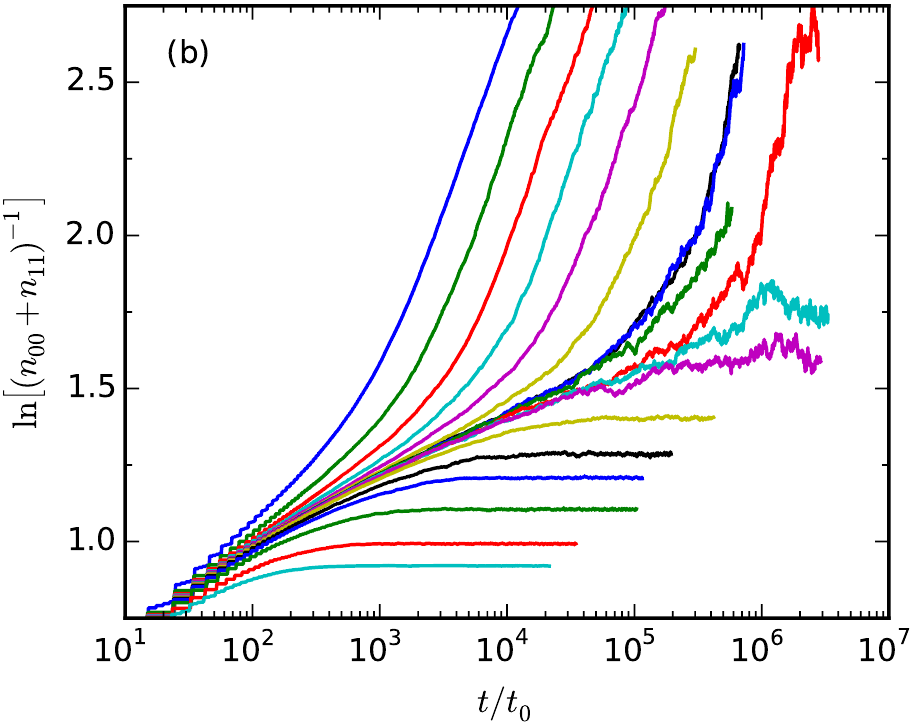}
	\caption{(Color online) Inverse of the concentration of active bonds vs.\
	time $t/t_0$ in (a) single- and (b) double-logarithmic plots.
	The results have been obtained from Monte Carlo simulations for
	$G=1$ and $L=8192$ and various values of $\Delta V/T$ (see text),	
	starting with an empty lattice, which corresponds to all bonds being
	active.}
    \label{fig:active_active_start}
\end{figure}

To check the DP2 universality further, we have investigated the time
evolution of the concentration $\rho_\mathrm{act}=n_{00}+n_{11}$ of active
sites, i.e., of empty-empty and occupied-occupied bonds, when we start from
a completely empty or completely occupied lattice, in which \emph{all} bonds are
active. In this case, the mean-field plus logarithmic form of the scaling
relation at criticality reads as
\begin{equation}
	n_{00}+n_{11} \sim \frac{1}{\ln t},
\label{eq:nn_mflog}
\end{equation}
whereas the power-law form is
\begin{equation}
	n_{00}+n_{11} \sim t^{-\alpha},
\label{eq:nn_powerlaw}
\end{equation}
where the best estimate is $\alpha=0.080(4)$ \cite{PhysRevE.83.011114}.
Our results are presented in Fig.\ \ref{fig:active_active_start}.
Figure \ref{fig:active_active_start}(a) [\ref{fig:active_active_start}(b)]
would show a straight line at the critical $\Delta V/T$
if $n_{00}+n_{11}$ satisfied mean-field plus logarithmic (power-law) scaling.
The data agree better with mean-field scaling with logarithmic corrections
for smaller $t$ but do not exclude a crossover to power-law scaling at larger
times.



In any case, while we cannot resolve the critical behavior,
the transition between regions $(4,1)$ and $(3,1)$
shows clear characteristics of the 2D DP2 universality class.
This is reasonable since a key feature of DP2 is the existence of
two symmetry-related absorbing states. Our model obviously has two
absorbing checkerboard states that are related by a lattice translation.
The DP2 character of the transition
supports our conclusion that the system in region $(4,1)$ is in the
active, uniform state, whereas in region $(3,1)$
it is in the absorbing, checkerboard state.
It is interesting that we find a DP2 transition in view of the
expectation that the non-equilibrium phase transitions of our model are
generically of Ising type \cite{PhysRevLett.97.236808, PhysRevLett.110.195301,
*PhysRevB.89.134310}. We conjecture that this is made possible by 
the fine tuning inherent in taking $\Delta V$ and $T$ to zero with
$\Delta V/T$ fixed.

The question arises as to
whether the system in the regions $(3,1)$, $(4,2)$, and
$(4,1)$ can be driven across the DP2 absorbing phase transition by varying $G$.
Increasing $G$ favors occupied over empty sites. We would thus expect it to
destabilize checkerboard order in favor of a uniform state with occupancy
close to unity. Since region $(4,1)$ is in the active phase even for $G=1$
we do not expect the active phase to be destroyed for any $G$. Indeed,
we have not found any sign of checkerboard order for $G$ values up to $10^8$.


\begin{figure}
\includegraphics[width=\columnwidth,keepaspectratio=true]{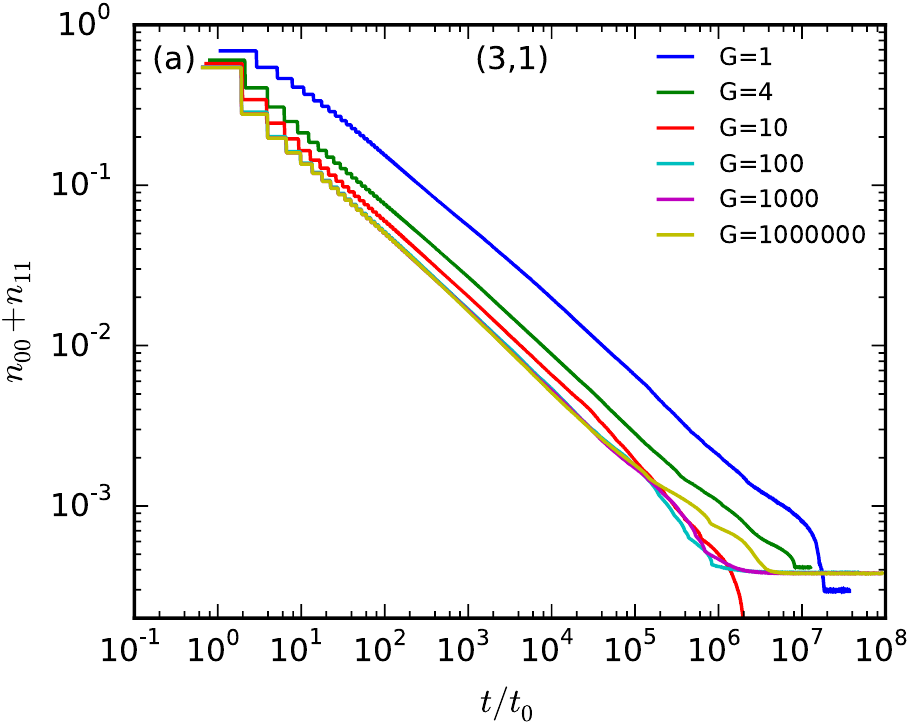}\\[2ex]
\includegraphics[width=\columnwidth,keepaspectratio=true]{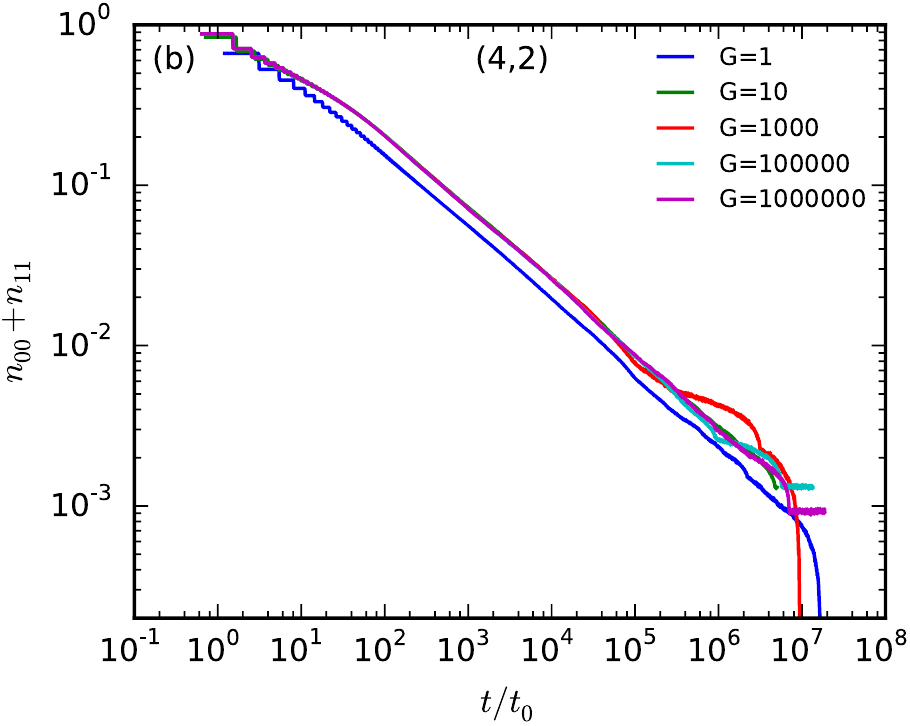}
	\caption{(Color online) Concentration of active sites for
	(a) region $(3,1)$ and (b) region $(4,2)$,
	from Monte Carlo simulations starting with a fully occupied lattice, for
	$L=8192$, $T/U_1 = 0$, and various values of $G$.}
	\label{fig:startHomogen_G}
\end{figure}

We now turn to the regions $(4,2)$ and $(3,1)$, in which a DP2 transition
as a function of $G$ might occur. We present simulation results for the
surviving concentration $n_{00}+n_{11}$ of active bonds for a fully occupied
(hence, active) starting configuration in Fig.\ \ref{fig:startHomogen_G}. 
There is no indication of a phase transition for
$G$ up to $10^6$. Note that the time evolutions shown in Fig.\
\ref{fig:startHomogen_G} sometimes get trapped in a seemingly stationary state
with nonzero active sites. These states consist of an even number of
straight domain walls of the checkerboard order spanning the
finite, periodic system. These domain walls are very
long lived under local updates since their annihilation requires them to
first deform and reconnect.
They clearly would not be possible in an infinite system.
Thus we conclude that the regions $(4,2)$ and $(3,1)$ are in the absorbing
phase for all $G$. In particular, the coexistence regime
found in the MFME phase diagram in Fig.\
\ref{fig:phase_diagramm_T0_Cluster_MF_ME} does not exist, only the
checkerboard blocked phase is stable here.
This general result is further supported by the power-law decay of
$n_{00}+n_{11}$ with time, in the limit of large $t$ but before finite-size
effects occur, with an exponent of approximately $1/2$.
This behavior was previously interpreted as being 
characteristic for the absorbing regime \cite{PhysRevLett.87.045701}.



\subsection{Phase diagram, occupation, and current}

\begin{figure}
	\includegraphics[width=\columnwidth,keepaspectratio=true]{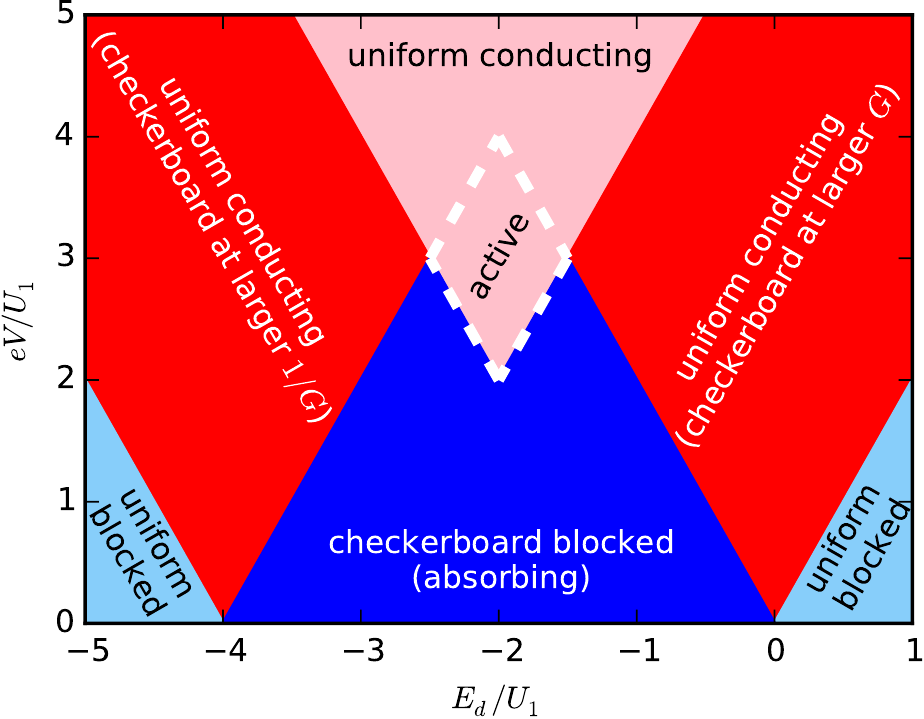}
	\caption{(Color online) Phase diagram obtained from Monte Carlo simulations for
	the monolayer with degeneracy $G=2$ and temperature $T=0$. Compare the
	MFME phase diagram in Fig.\ \ref{fig:phase_diagramm_T0_Cluster_MF_ME}.
    In the region marked ``active,'' the layer is in the uniform conducting
    state even though the two states with perfect checkerboard charge order
    (and no current) are absorbing.}
	\label{fig:phase_diagramm_T0_MC}
\end{figure}

\begin{figure}
	\includegraphics[width=\columnwidth, keepaspectratio=true]{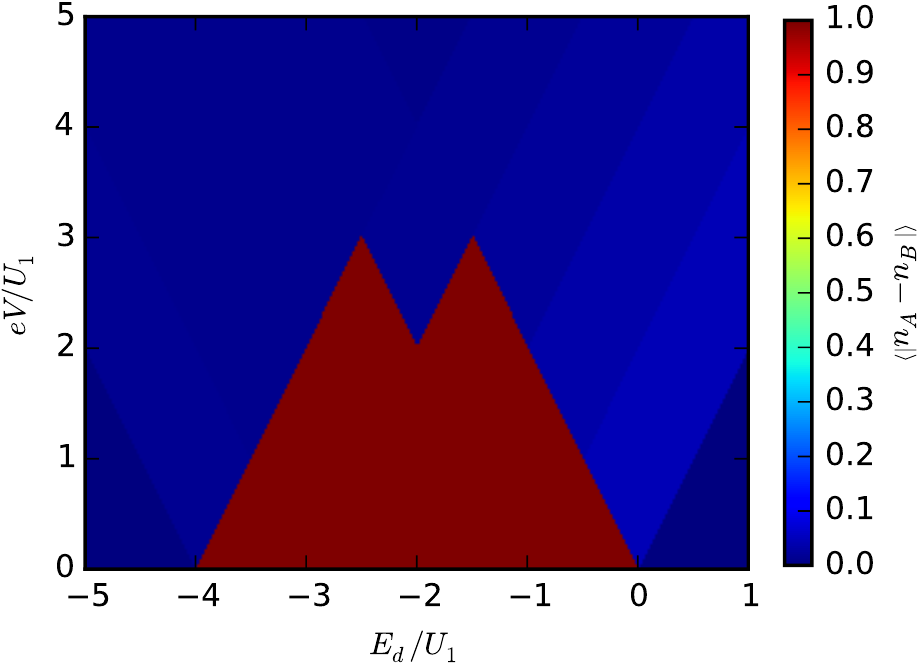}
	\caption{(Color online) Monte Carlo results for the 
	average imbalance between the occupation numbers per site on the two
	sublattices, $\expect{\left| n_A-n_B \right|}$, for $T=0$, $L=64$,
	and $G=2$. In the central region, the system ends up in one of
	the two	absorbing checkerboard-ordered states so that the 
	order parameter is exactly $1/2$.}
	\label{fig:T0_MC_order}
\end{figure}

\begin{figure}
	\includegraphics[width=\columnwidth, keepaspectratio=true]{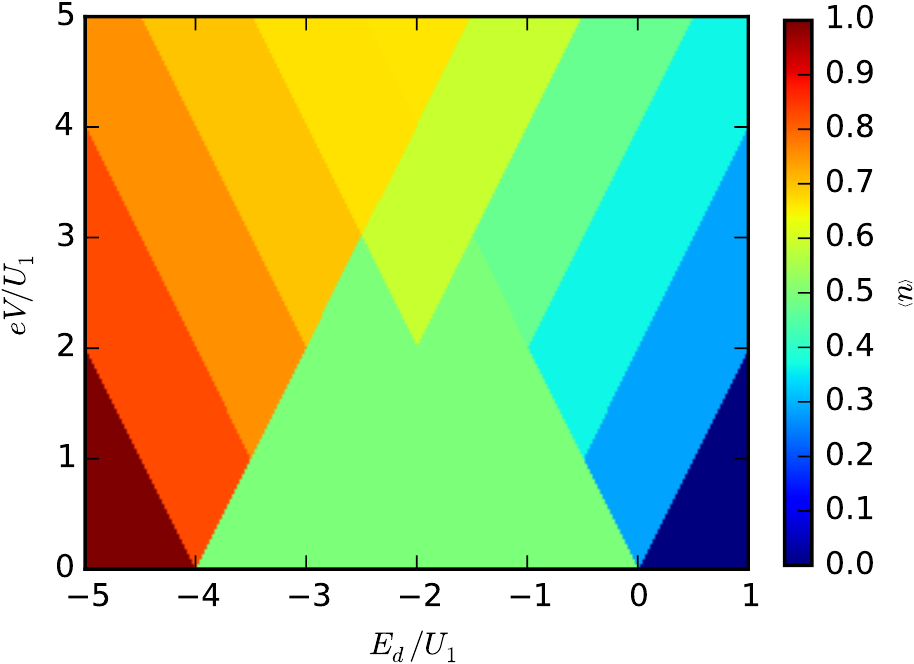}
	\caption{(Color online) Monte Carlo results for the
	average occupation per site, $\expect{n}$,
	for the same parameters as in Fig.\ \ref{fig:T0_MC_order}.
	The occupation is $0$  and $1$ in the regions $(0,0)$
	and $(5,5)$, respectively, and $1/2$ in the checkerboard-ordered region.
	The average occupation assumes a non-universal value in the other regions,
	which grows  for smaller on-site energy $E_d$ and tends toward
	$G/(G+1)=2/3$ for increasing bias voltage $V$.}
	\label{fig:T0_MC_occupation}
\end{figure}

\begin{figure}
	\includegraphics[width=\columnwidth, keepaspectratio=true]{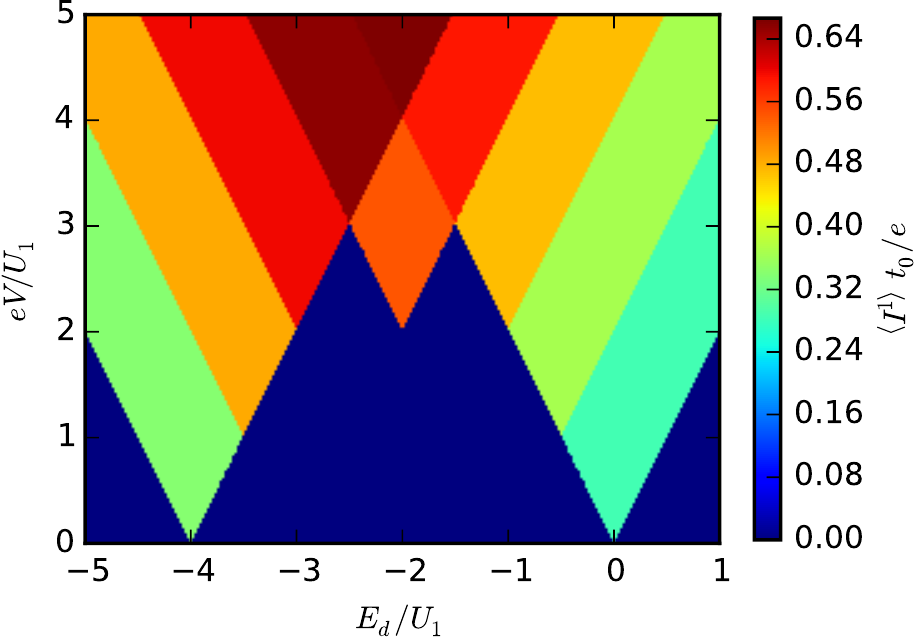}
	\caption{(Color online) Monte Carlo results for the average current
	per site through the lead $\alpha=1$, Eq.\ \eqref{eq:current},
	for the same parameters as in Fig.\ \ref{fig:T0_MC_order}. The
	current vanishes in the absorbing phases and grows with the number of
	transitions in the bias window, i.e., with the  bias voltage $V$.}
	\label{fig:T0_MC_current}
\end{figure}

In the previous subsections, we have discussed the stationary state 
at $T=0$ for all the regions in Fig.\ 
\ref{fig:phase_diagramm_T0_energy_crossings}. The results are summarized in
the phase diagram in Fig.\ \ref{fig:phase_diagramm_T0_MC},
which partially anticipates results for the current that are presented below.
We mention in passing that, although the simulations are not
restricted to uniform and checkerboard-ordered phases, we did not find any
other type of charge order.
One of the most intriguing results is the possibility of
\emph{bias-induced charge order}: consider for example an on-site
energy $E_d>0$. If the degeneracy $G$
is sufficiently large and we increase the bias voltage
$V$ starting from zero, the system is initially in a uniform blocked phase
with all sites empty. But at $eV=2E_d$ it enters a conducting phase with
checkerboard charge order, see Sec.\ \ref{sec.results.zero.CCG}.
At a higher bias, there is a second transition
towards a uniform conducting phase.

While the applied bias voltage $V$ is easily varied, this is not the
case for the on-site energy $E_d$. One could, however, prepare
series of devices with different on-site energies. In molecular monolayers, 
one can tune $E_d$ by interchanging side groups, as studied experimentally in Ref.\
\cite{NanoLett.7.502} and theoretically in Ref.\ \cite{JChemPhys.128.111103}.
Tuning $E_d$ \textit{in situ} is more difficult. Note that tuning $E_d$ and $V$ is
equivalent to changing the potential drops between the molecules
and the two electrodes independently. This could be done by
asymmetrically changing the molecule-electrode distances: if the molecules are
covalently bound to one electrode but only
van\ der\ Waals-coupled to the other, changing the electrode-electrode
separation would have the desired effect. This would of course also change the
tunneling amplitudes.

We now present simulation results for the most relevant observables
in the stationary state for $G=2$, corresponding to
pure spin degeneracy of the occupied sites.
The average imbalance between the occupation numbers per site on the two
sublattices, $\expect{\left| n_A-n_B \right|}$, is shown in
Fig.\ \ref{fig:T0_MC_order}, the average occupation in
Fig.\ \ref{fig:T0_MC_occupation}, and the average current
in Fig.\ \ref{fig:T0_MC_current}.
As discussed before, there is a uniform fully occupied phase for $E_d/U_1 < -4$,
a corresponding uniform completely empty phase for $E_d/U_1 > 0$, and a perfectly
checkerboard-ordered phase in between. All of them are blocking the current due 
to the absence of allowed transitions in the bias window. For increasing
bias, the monolayer will eventually become conducting and disordered as more and
more transitions enter the bias window. 
Note that $\expect{\left| n_A-n_B \right|}$ is not identically zero in
the conducting phases since the nonzero current implies fluctuations in the
occupation numbers. All three quantities plotted in Figs.\
\ref{fig:T0_MC_order}--\ref{fig:T0_MC_current} clearly show a double-peaked
blocked region resulting from the appearance of an active phase in region
$(4,1)$ (see Fig.\ \ref{fig:phase_diagramm_T0_energy_crossings}).
The current and the average occupation
approach and finally reach the non-interacting single-site limit as
the number of transitions in the bias window is increased.
The observables are asymmetric in the on-site energy
relative to $E_d/U_1 = -2$ since the degeneracy $G=2$ breaks particle-hole
symmetry.

It would be desirable to verify the checkerboard charge order
experimentally. The charge order implies the presence of two populations of molecules
or quantum dots with
distinct average charge, each comprising $50\%$ of the monolayer. In the blocked
phase, charge fluctuations are suppressed. For molecular layers, the resulting
equal distribution of charge states could be seen by optical spectroscopy,
for example, in reflection geometry with a transparent conductor as top electrode.
On the other hand, in the checkerboard conducting phase, the occupation number of
one population fluctuates due to the current, whereas the other is essentially fixed.
Spectroscopy should thus see two charge states but with different probabilities.
For a layer of metallic nanoparticles, one could similarly try to observe the
presence of two populations with distinct surface-plasmon frequencies.
A more challenging idea is to observe the diffraction pattern due to the
diffraction grating formed by the charge density wave.

\section{Results for nonzero temperatures}
\label{sec.results.nonzero}

In this section, we present results for nonzero temperatures. While this is
clearly required for comparison with experiments, the
case of $T>0$ is less interesting from the point of view of statistical physics.
The nontrivial DP2 transition found for $T=0$ in Sec.\ \ref{sec.results.zero.absorb}
relies on the perfect checkerboard states being absorbing, which is no longer
the case for $T>0$. Hence, we expect all (equilibrium and non-equilibrium) 
phase transitions to be in the 2D Ising universality class
\cite{PhysRevLett.97.236808}. This also holds for those transitions that
were trivially discontinuous for $T=0$ due to the jump in the Fermi function.
At $T>0$, the Fermi functions and consequently the transition 
rates $R_{i\to f}$ in Eq.\ \eqref{eq:seq_fullrates} are continuous 
functions of the parameters $E_d/U_1$, $eV/U_1$, and $T/U_1$.

\begin{figure}
	\includegraphics[width=\columnwidth, keepaspectratio=true]{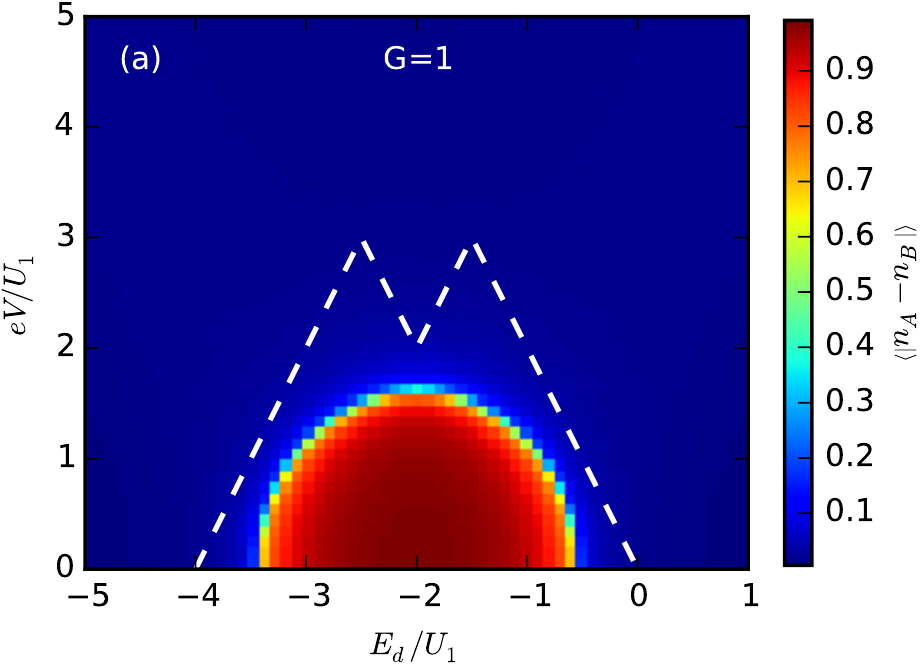}
	\includegraphics[width=\columnwidth, keepaspectratio=true]{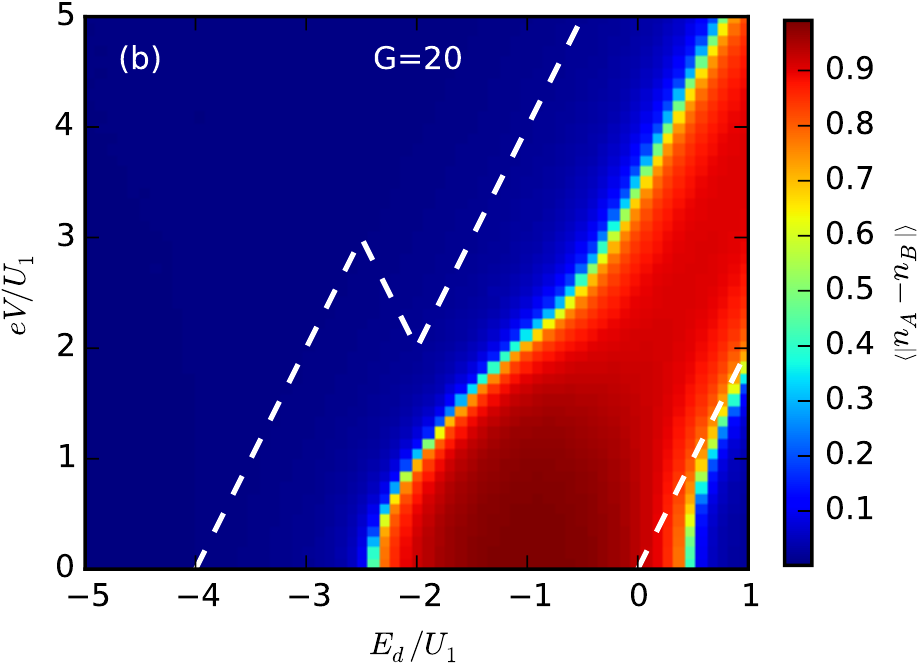}
	\caption{(Color online) Checkerboard order parameter $\expect{
	\left| n_A-n_B \right| }$ from Monte Carlo simulations for
	$T/U_1=0.35$ and $L=64$. The degeneracy is (a) $G=1$ and
	(b) $G=20$. The dashed lines denote the boundary between
	the checkerboard ordered and uniform phases for $T=0$.
	Note that even though the temperature is a sizable fraction of the zero-field
	critical temperature of the Ising model, the large
	degeneracy in panel (b) still stabilizes
	the checkerboard conducting phase.}
	\label{fig:phase_diagramm_T_MC}
\end{figure}

Figure \ref{fig:phase_diagramm_T_MC} shows results for the checkerboard
order parameter for a temperature of approximately two thirds of the
Ising critical temperature $T_c \approx 0.567\,U_1$ and two values of $G$,
the particle-hole symmetric case $G=1$ and the large degeneracy $G=20$.
The observed shrinking of the regime with checkerboard order compared to $T=0$
is of course expected since higher temperatures allow additional tunneling
processes that tend to destabilize charge order.
For $G=20$, the ordered regime is also shifted to larger $E_d$, in accordance
with the effective on-site energy $E_d-T\ln G$ (see Sec.\
\ref{sec.results.zero.CCG}). Interestingly, for this large value of $G$, the
checkerboard conducting phase is found to be rather robust against thermal
fluctuations.

\begin{figure}
\includegraphics[width=\columnwidth,keepaspectratio=true]{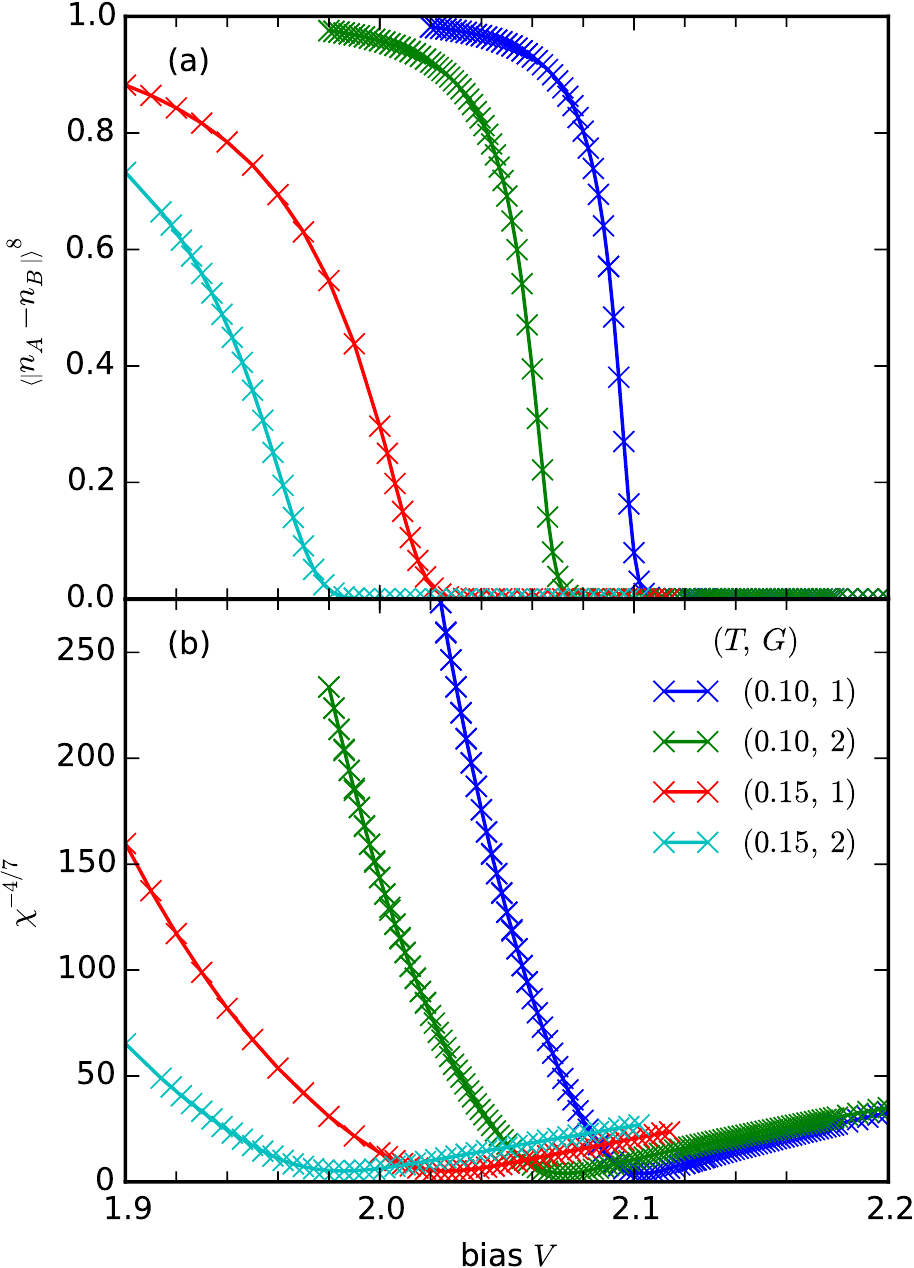}
	\caption{(Color online) Monte Carlo results for (a) the checkerboard order
	parameter $\expect{\left| n_A-n_B \right| }$ to the power $8$ and
	(b) the corresponding susceptibility to the power $-4/7$
	as functions of the bias voltage $V$ for various values of $G$ and $T$.
	The remaining parameters are $E_d/U_1=2.2$ and $L=64$.
	The results are consistent with the 2D Ising universality class.}
	\label{fig:phasTransV_Chi47_op8}
\end{figure}

The character of the phase transitions can obviously not be inferred
from Fig.\ \ref{fig:phase_diagramm_T_MC}. We exemplarily consider the
transition between checkerboard and uniform phases driven by the bias voltage
$V$ at a fixed on-site energy $E_d$ and several values of $T$ and $G$. 
As in Sec.\ \ref{sec.results.zero.CCG}, we plot the checkerboard order
parameter to the power $8$ and the corresponding susceptibility to the
power $-4/7$ in Fig.\ \ref{fig:phasTransV_Chi47_op8}. The results are
consistent with the 2D Ising universality class~\cite{PhysRevLett.97.236808,
pathria2011statistical_mab216023733, PhysRevLett.110.195301, *PhysRevB.89.134310}.

\begin{figure}
\includegraphics[width=\columnwidth,keepaspectratio=true]{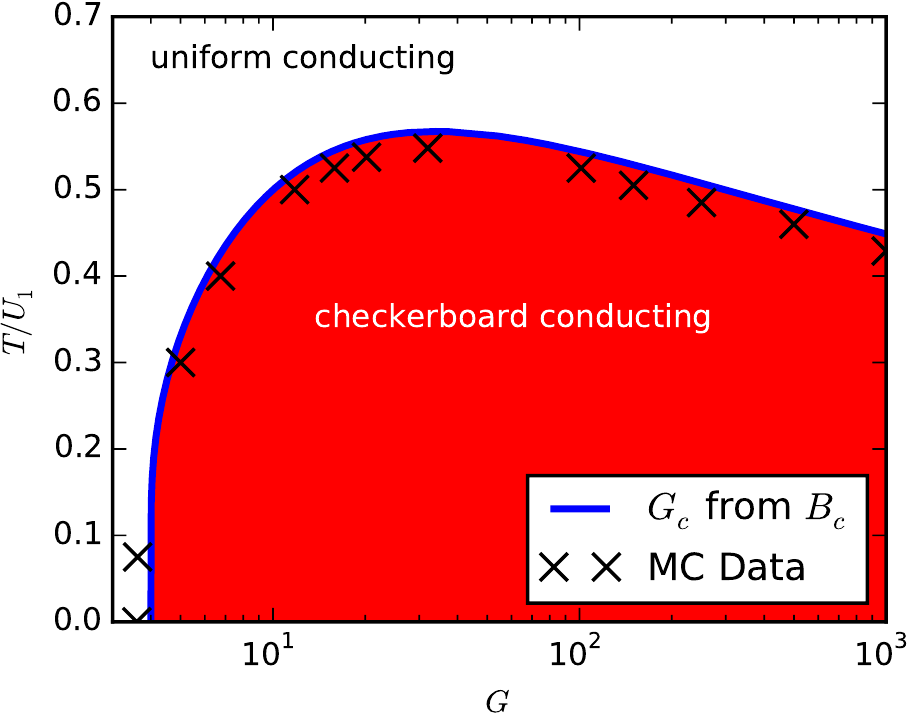}
	\caption{(Color online) Phase diagram in the \textit{G}-\textit{T} plane
	for $eV/U_1 = E_d / U_1 = 0$, i.e., in region $(1,0)$. The crosses denote
	the location of the phase
	transition determined from the maximum of the susceptibility $\chi$
	obtained from Monte Carlo simulations for $L=64$.
	The continuous line is based on the conjecture for the
	critical magnetic field in Ref.~\cite{doi.10.1007.BF01325537}.}
	\label{fig:CCG_TG_phase_diagram}
\end{figure}

We finally turn to the checkerboard conducting regions, specifically the
region $(1,0)$, at nonzero temperatures. We found in Sec.\
\ref{sec.results.zero.CCG} that at $T=0$ there is a transition between uniform
and checkerboard-ordered conducting states as the degeneracy $G$ is increased.
For this, it is important that a checkerboard state cannot be destroyed by
electrons tunneling in to create occupied defects in
the empty sublattice since
the corresponding rate vanishes. This is no longer true for $T>0$; for low
temperatures there is now an exponentially small rate for creating such
occupied defects. However, this in-tunneling rate contains a factor of $G$ so
that for increasing $G$ the checkerboard state should eventually be destabilized
in favor of a uniform state with occupancy close to unity.
The same conclusion is reached by considering the effective on-site energy
$E_d-T\ln G$ in the equivalent Ising model.
This expectation is indeed borne out by the results presented
in Fig.\ \ref{fig:CCG_TG_phase_diagram}.
We find a reentrant transition to the uniform conducting state at large
$G$, which shifts to smaller $G$ for increasing temperature. On the other hand,
the transition to checkerboard order at lower $G$ shifts upwards with
temperature. Both trends are consistent with the expectation that higher
temperatures disfavor ordering. Above
the corresponding critical temperature of the 2D Ising model,
checkerboard order does not exist for any $G$.
Since Fig.\ \ref{fig:CCG_TG_phase_diagram} shows results for $V=0$, i.e.,
in equilibrium, the critical temperature versus $G$ should map to the critical
temperature versus applied magnetic field for the Ising model. Our simulation
results indeed agree well with the conjecture of M\"uller-Hartmann and
Zittartz~\cite{doi.10.1007.BF01325537}.

\section{Summary and conclusions}
\label{sec:summary}

In summary, we have studied a square lattice of quantum dots or
molecules under a bias voltage applied
perpendicular to the layer. We have assumed 
infinite on-site repulsion and finite nearest-neighbor
Coulomb interaction within the layer, as well as vanishing intralayer
hopping and weak monolayer-electrode hopping. The indirect hopping
from a site in the monolayer to one of the electrodes and further to a
\emph{different} site of the monolayer is assumed to be fully incoherent,
which is the case in the limits of short Fermi wavelength or strong
disorder in the electrodes. By employing Monte Carlo simulations, we
avoid mean-field approximations.
The interactions lead to the appearance of charge-density-wave
phases. Apart from the charge order, the main quantity of interest is the
current perpendicular to the layer.

The resulting zero-temperature phase diagram, Fig.\
\ref{fig:phase_diagramm_T0_MC}, shows
blocked phases with vanishing current and zero, single,
or checkerboard-ordered occupancy. The latter can be understood as a
Coulomb-blockade state induced by the nearest-neighbor repulsion.
These phases are connected to an equilibrium Ising model at $V=0$.
At larger bias voltages, they give way to conducting phases. Interestingly,
it is possible for these phases to possess checkerboard charge order.
This requires a high degeneracy $G\gtrsim 3.6$
of the occupied single-site states, which could
be realized by combining charge and orbital degeneracies. The transition
between the uniform conducting phase
and the checkerboard conducting phase as a function of $G$
is in the 2D Ising universality class.
The checkerboard conducting phase only exists out of equilibrium.
In a large parameter range we find a transition from a uniform blocked phase to
a checkerboard conducting phase at a finite critical bias voltage. This
constitutes an interesting case of bias- or current-induced charge order.
Furthermore, there is a region at finite bias voltage for which the two
symmetry-related blocked checkerboard states are absorbing but the stationary
state is nevertheless conducting and uniform. 
The presence of this active phase is evident in the current-voltage
characteristics. It is interesting that such an active phase could
be realized in a monolayer under bias, as there are not many experimental
realizations. By judiciously taking the limit $T\to 0$, we determine
the phase transition between the absorbing and
active phases to be in the 2D DP2 universality class.


The features found at $T=0$ are robust for small nonzero temperatures,
except that absorbing states no longer exist for $T>0$ and that, as a
consequence, the
absorbing-to-active phase transition transforms into a 2D Ising transition
between checkerboard blocked and uniform conducting phases. Apart from this
change, the ordered phases shrink and, for degeneracies $G>1$, shift to higher
on-site energies for increasing temperature.

It would be desirable to extend the underlying dynamics to
contain coherences as well as
higher-order tunneling processes  such as cotunneling  and
pair tunneling. Even in the absence of intralayer hopping,
tunneling via the electrodes can induce coherences
between eigenstates of the
local particle numbers, i.e., delocalized charges in the monolayer.
Intralayer hopping of course also favors delocalization in the
monolayer. Higher-order processes and coherences are required
for a study of Kondo-type effects in tunneling through the layer.
Coherent hopping, be it direct or indirect through the electrodes,
would turn the system into a much more difficult
extended Hubbard model out of equilibrium. This would call for
non-equilibrium \emph{quantum} Monte Carlo simulations, which would
suffer from the sign problem. On the other hand, a higher-order
MFME including coherences seems feasible. In any case, even the quasi-classical
model considered here should be valuable for further studies. On the one hand,
comparison with experiments, for example on rolled-up structures, calls for
a realistic description of electronic, spin, and vibrational degrees of
freedom of molecular layers. On the other,
further studies of the considered model might help to constrain
the critical behavior of the 2D DP2 universality class.


\begin{acknowledgments}
We would like to thank S. Diehl, J. Marino, A. Rubio, T. Vojta,
and A. Wacker for useful discussions. Financial support by the 
Deutsche Forschungsgemeinschaft, through Research Unit
FOR 1154, \textit{Towards Molecular Spintronics}, is gratefully
acknowledged. C.\,T. also acknowledges support by the Deutsche
Forschungsgemeinschaft through Collaborative Research Center SFB 1143.
\end{acknowledgments}

\bibliography{ludwig}

\end{document}